\documentclass[aps,prl,reprint,superscriptaddress,nofootinbib,longbibliography,floatfix]{revtex4-2}

\usepackage{amsmath,amssymb,mathtools,bm,mathrsfs}
\usepackage{microtype}
\usepackage[colorlinks=true,citecolor=blue,linkcolor=blue,urlcolor=blue]{hyperref}

\newcommand{\Res}{\operatorname{Res}}
\newcommand{\I}{\mathcal I}
\newcommand{\dd}{\mathrm d}

\newcommand{\ABJM}{\mathrm{ABJM}}
\newcommand{\BMN}{\mathrm{BMN}}

\newcommand{\Hom}{\operatorname{Hom}}

\begin{document}

\title{ABJM/BMN Index Matching in the Plane-Wave Limit}

\author{Chi-Ming Chang}
\email[Contact author: ]{cmchang@tsinghua.edu.cn}
\affiliation{Yau Mathematical Sciences Center (YMSC), Tsinghua University, Beijing, China}
\affiliation{Beijing Institute of Mathematical Sciences and Applications (BIMSA), Beijing, China}
\affiliation{Peng Huanwu Center for Fundamental Theory, Hefei, Anhui, China}

\author{Kangning Liu}
\email[Contact author: ]{lkn22@mails.tsinghua.edu.cn}
\affiliation{Yau Mathematical Sciences Center (YMSC), Tsinghua University, Beijing, China}
\affiliation{Department of Mathematical Sciences, Tsinghua University, Beijing, China}
\affiliation{Jefferson Physical Laboratory, Harvard University, Cambridge, Massachusetts 02138, USA}

\date{\today}

\begin{abstract}
We establish coefficientwise equality between the refined supersymmetric indices of ABJM theory and the BMN matrix model in the plane-wave limit at every fixed positive Chern--Simons level $k$. For multi-membrane families with fixed multiplicities---and hence fixed finite ABJM gauge rank---the equality holds as all longitudinal momenta tend to infinity with their pairwise differences fixed. Scalar-dressed equal-flux ABJM monopoles correspond to BMN vacua of fuzzy-sphere membranes: multiplicities count coincident membranes, whereas flux magnitudes give the momentum per membrane along the M-theory circle. At the index level, the associated Higgs residues yield a diagonal-holonomy integral for the ABJM plane-wave contribution, whose monopole-harmonic kernel agrees with the BMN matrix-harmonic kernel up to the finite BMN upper-spin cutoff. At finite momentum, the full index difference splits into the non-plane-wave remainder and the harmonic tail above this cutoff. Increasing $k$ shifts the former to higher fugacity grade without changing the surviving kernel or the cutoff contribution. This gives a coefficientwise, multi-membrane extension of the earlier single-membrane index relation.

\end{abstract}

\maketitle

The plane-wave limit of the $AdS_4/CFT_3$ correspondence provides a tractable setting in which a large-charge sector of ABJM theory can be compared with the BMN matrix model.  On the bulk side, taking the Penrose limit of M-theory on $AdS_4\times S^7/\mathbb Z_k$ about a trajectory with large momentum along the M-theory circle isolates excitations of finite light-cone energy and yields the maximally supersymmetric eleven-dimensional plane wave \cite{Blau:2002dy,Kovacs:2013una}.  On the ABJM side, the dual sector is represented in radial quantization by scalar-dressed monopole states whose monopole charges encode this circle momentum \cite{SheikhJabbari:2009ns,Ezhuthachan:2011dx}.  The BMN matrix model, in turn, describes M-theory on the same plane-wave background in discrete light-cone quantization \cite{Banks:1996vh,Berenstein:2002jq,Dasgupta:2002hx}.  Together, these observations motivate testing the correspondence through a direct comparison of the ABJM and BMN supersymmetric indices.

The two descriptions package the same membrane degrees of freedom differently.  The BMN model is the maximally supersymmetric mass deformation of the $U(\mathcal N)$ BFSS matrix quantum mechanics \cite{Banks:1996vh,Berenstein:2002jq}.  Its fields include nine Hermitian matrix coordinates $X^I(t)$, $I=1,\ldots,9$, their fermionic partners, and a one-dimensional gauge connection.  The mass and Myers couplings single out a triplet $X^a$, $a=1,2,3$, which can form fuzzy spheres; the remaining six coordinates vanish in the vacua \cite{Berenstein:2002jq,Dasgupta:2002hx}.  ABJM theory is a three-dimensional Chern--Simons matter theory with gauge group $U(r)_k\times U(r)_{-k}$, gauge connections $A_\mu$ and $\widehat A_\mu$, four complex bifundamental scalars $Y^A$, $A=1,\ldots,4$, and their fermionic partners \cite{Aharony:2008ug}. 

Previous work established the vacuum correspondence between scalar-dressed ABJM monopoles and BMN fuzzy-sphere membranes \cite{SheikhJabbari:2009ns,Ezhuthachan:2011dx,Kovacs:2013una}. Berenstein and Park studied quadratic fluctuations about half-BPS ABJM monopole backgrounds \cite{Berenstein:2009sa}. In the large-$r$ semiclassical Penrose regime $r^{1/3}\ll J\ll r^{1/2}$ with $k=O(1)$, Ref.~\cite{Kovacs:2013una} matched the complete tree-level single-membrane BMN spectrum to the leading large-flux ABJM spectrum and compared selected two-membrane modes. To our knowledge, an analogous full-spectrum comparison for general multi-membrane backgrounds remains open.

Protected indices provide a complementary test of the correspondence. In the special case of ABJM rank $r=1$ and level $k=1$, Ref.~\cite{Chang:2024lkw} found that, after fugacity specialization, the ABJM index summed over monopole sectors agrees with the large-$\mathcal N$ index about an irreducible BMN vacuum up to a divergent factor. Here we extend the comparison to fixed monopole sectors at arbitrary fixed finite ABJM rank and every fixed positive integer $k$.

Let $\I_{\ABJM,k}^{\vec m,\widetilde{\vec m}}(a,b,c)$ and $\I_{\BMN}^{\mathfrak p}(a,b,c)$ denote, respectively, the refined supersymmetric index in a fixed ABJM Goddard--Nuyts--Olive (GNO) sector $(\vec m,\widetilde{\vec m})$ and the index about a BMN vacuum labeled by a partition $\mathfrak p$ of the BMN rank $\mathcal N$, both expressed in the common fugacities $a,b,c$. We restrict throughout to sectors with equal left and right GNO charges, $\vec m=\widetilde{\vec m}$, relevant to the BMN vacuum dictionary. Their trace definitions and charge map are given in the Supplemental Material \cite{Kim:2009wb,Chang:2024lkw,SupplementalMaterial}.

The main result of this paper is, for every fixed $k\in\mathbb Z_{>0}$,
\begin{equation}
 \boxed{
 \lim_{\Lambda\to\infty}
 \I_{\ABJM,k}^{\vec m_\Lambda,\widetilde{\vec m}_\Lambda}(a,b,c)
 =
 \lim_{\Lambda\to\infty}
 \I_{\BMN}^{\mathfrak p_\Lambda}(a,b,c).}
 \label{eq:main-result}
\end{equation}
The equality in Eq.~\eqref{eq:main-result} is coefficientwise as a formal series in nonnegative half-powers of $a,b,c$. Explicitly, for every fixed $r_i\in\frac12\mathbb Z_{\geq0}$, the coefficients of $a^{r_1}b^{r_2}c^{r_3}$ in the two finite-$\Lambda$ indices agree for all sufficiently large $\Lambda$. Equivalently, for every fixed $D$, all coefficients with $r_1+r_2+r_3\leq D$ eventually agree.

To define the large-momentum family appearing in Eq.~\eqref{eq:main-result}, choose the fixed number $s$ of distinct block sizes, their fixed positive multiplicities $n_\alpha$, and pairwise-distinct fixed integer offsets $\nu_\alpha$.  For $\Lambda\in\mathbb Z$ large enough that $\Lambda+\nu_\alpha>0$ for every $\alpha$, set
\begin{equation}
 \begin{aligned}
 N_\alpha&\equiv\Lambda+\nu_\alpha,
 \qquad \alpha=1,\ldots,s,\\
 \vec m_\Lambda=\widetilde{\vec m}_\Lambda
 &=\big((-N_\alpha)^{n_\alpha}\big)_{\alpha=1}^{s},\\
 \mathfrak p_\Lambda
 &=\big\{(n_\alpha,N_\alpha)\big\}_{\alpha=1}^{s}.
 \end{aligned}
 \label{eq:limit-shorthand}
\end{equation}
Equation~\eqref{eq:main-result} takes $\Lambda\to\infty$ at fixed $n_\alpha,\nu_\alpha$, so the ABJM rank $r=\sum_\alpha n_\alpha$ stays finite while $\mathcal N$ grows.

We now explain the physical background. A BMN vacuum is labeled by a partition of the matrix size,
\begin{equation}
 \mathcal N=\sum_{\alpha=1}^{s}n_\alpha N_\alpha,
 \qquad
 X^a_{\rm vac}=\frac{\mu}{3}\bigoplus_{\alpha=1}^{s}
 \left(\mathbf 1_{n_\alpha}\otimes J^a_{(N_\alpha)}\right),
 \label{eq:bmn-vacuum}
\end{equation}
where $\mu$ is the plane-wave mass and $J^a_{(N_\alpha)}$ are the generators of the $N_\alpha$-dimensional irreducible $SU(2)$ representation. Each irreducible block is a supersymmetric fuzzy-sphere saddle: the matrix regularization of a spherical M2-brane in the eleven-dimensional plane wave \cite{Berenstein:2002jq,Dasgupta:2002hx}. For $AdS_4\times S^7/\mathbb Z_k$, let $\varphi$ denote the covering-space Hopf-fiber angle, with period $2\pi$.  The orbifold identifies $\varphi\sim\varphi+2\pi/k$. We denote the angular momentum on the covering space as $J=i\partial_\varphi$. In the pp-wave matrix description, the longitudinal quotient is implemented through momentum quantization: each irreducible block carries covering-space charge $J_\alpha^{\mathrm{block}}=kN_\alpha$ and quotient-circle momentum $P_\alpha^{\mathrm{block}}=N_\alpha$, while $n_\alpha$ counts coincident membranes of that size \cite{Kovacs:2013una}. The unbroken gauge group of this BMN vacuum is $\prod_\alpha U(n_\alpha)$. The corresponding ABJM sector has gauge rank $r=\sum_\alpha n_\alpha$ and equal left and right GNO charges with $n_\alpha$ copies of $-N_\alpha$. Thus the BMN block multiplicity $n_\alpha$ becomes the multiplicity of an ABJM monopole flux, while the block size $N_\alpha$ becomes its flux magnitude and, as shown below, the quotient M-circle momentum per membrane. The monopole is dressed by a rotating bifundamental scalar zero mode whose condensate breaks the two gauge factors to the diagonal subgroup, matching the unbroken gauge group of the BMN vacuum. 
The vacuum in Eq.~\eqref{eq:bmn-vacuum} is therefore the finite-$\Lambda$ member of the family defined in Eq.~\eqref{eq:limit-shorthand}.

The coefficientwise agreement in Eq.~\eqref{eq:main-result} occurs only in the $\Lambda\to\infty$ limit. For the finite-$\Lambda$ member of this family, abbreviate $\I_{\ABJM,k;\Lambda}\equiv\I_{\ABJM,k}^{\vec m_\Lambda,\widetilde{\vec m}_\Lambda}(a,b,c)$ and $\I_{\BMN;\Lambda}\equiv\I_{\BMN}^{\mathfrak p_\Lambda}(a,b,c)$. Let $\I_{\ABJM,k;\Lambda}^{\rm pw}$ denote the contribution to $\I_{\ABJM,k;\Lambda}$ obtained by retaining only the plane-wave Higgs residues; its explicit residue sum is given in Eq.~\eqref{eq:pw-index-definition}. The two finite-$\Lambda$ sector indices then differ by two corrections whose large-$\Lambda$ behavior can be analyzed independently:
\begin{align}
 \I_{\ABJM,k;\Lambda}-\I_{\BMN;\Lambda}
 ={}&\underbrace{(\I_{\ABJM,k;\Lambda}-\I_{\ABJM,k;\Lambda}^{\rm pw})}_{\text{non-plane-wave remainder}}
 \nonumber\\[-2pt]
 &+\underbrace{(\I_{\ABJM,k;\Lambda}^{\rm pw}-\I_{\BMN;\Lambda})}_{\text{harmonic tail}}.
 \label{eq:two-corrections-main}
\end{align}
The first correction measures the remainder after selecting the plane-wave Higgs residues of the fixed-GNO ABJM index, which involves two sets of gauge holonomies, $z_p$ and $w_p$. This remainder vanishes below a fugacity grade growing linearly with $k\Lambda$. The second correction comes from the fuzzy-sphere cutoff: at finite matrix size, the BMN harmonics truncate the continuum ABJM monopole-harmonic tower at a finite upper spin. This correction is independent of $k$.

\emph{ABJM half-BPS solutions.---}
 To make the ABJM side of the vacuum dictionary explicit and identify the Higgs residues entering $\I_{\ABJM,k;\Lambda}^{\rm pw}$, we summarize the relevant scalar-dressed half-BPS monopole saddles on $\mathbb R_t\times S^2_L$, where $L$ is the sphere radius \cite{SheikhJabbari:2009ns,Ezhuthachan:2011dx}. Write $Y^A=u^A Z$, with $\bar u_Au^A=1$, and let $F,\widehat F$ be the two gauge-field strengths, with GNO flux matrices $M=(2\pi)^{-1}\int_{S^2}F$ and $\widehat M=(2\pi)^{-1}\int_{S^2}\widehat F$. The BPS equations and the flux-intertwining condition $MZ=Z\widehat M$ are derived in the Supplemental Material \cite{SupplementalMaterial}. On this branch, the Chern--Simons Gauss laws reduce to the moment-map constraints
\begin{equation}
 kM=-4\pi L\,ZZ^\dagger,
 \qquad
 k\widehat M=-4\pi L\,Z^\dagger Z .
 \label{eq:gauss-main}
\end{equation}
Since $ZZ^\dagger$ and $Z^\dagger Z$ have the same nonzero eigenvalues and the two gauge groups have equal rank, a right Weyl permutation can therefore be used to set $M=\widehat M$. In the negative-flux convention, the most general half-BPS solution associated with Eq.~\eqref{eq:bmn-vacuum} is \cite{Ezhuthachan:2011dx}
\begin{align}
 M=\widehat M&=-\bigoplus_{\alpha=1}^{s}N_\alpha\mathbf 1_{n_\alpha},
 \nonumber\\[-2pt]
 Y^A_{(0)}(t)&=u^A e^{it/(2L)}\bigoplus_{\alpha=1}^{s}
 v_\alpha U_\alpha,
 \qquad
 |v_\alpha|^2=\frac{kN_\alpha}{4\pi L},
 \label{eq:abjm-saddle}
\end{align}
Equivalently, the left and right GNO eigenvalues are $m_{\alpha,p}=\widetilde m_{\alpha,p}=-N_\alpha$, $p=1,\ldots,n_\alpha$, reproducing the sector vector in Eq.~\eqref{eq:limit-shorthand}.
The matrices $U_\alpha\in U(n_\alpha)$ are removed by blockwise left--right gauge transformations. We set $U_\alpha=\mathbf 1_{n_\alpha}$ and use a global $SU(4)$ rotation to choose $u^A=\delta^{A4}$. The resulting condensate is invariant only under $g_{L,\alpha}=g_{R,\alpha}$ and breaks the gauge factors to the diagonal subgroup:
\begin{equation}
\prod_\alpha\left[U(n_\alpha)_L\times U(n_\alpha)_R\right]
 \longrightarrow \prod_\alpha U(n_\alpha)_{\rm diag}.
 \label{eq:classical-locking}
\end{equation}
This residual group is isomorphic to the unbroken gauge group of the corresponding BMN vacuum. For the $\alpha$th stack, the Gauss law fixes the total covering-space scalar charge and quotient M-circle momentum to be
$J_\alpha^{\mathrm{stack}}=n_\alpha J_\alpha^{\mathrm{block}}=k n_\alpha N_\alpha$ and
$P_\alpha^{\mathrm{stack}}=n_\alpha P_\alpha^{\mathrm{block}}=n_\alpha N_\alpha$, respectively.  Hence the total quotient M-circle momentum is $\sum_\alpha P_\alpha^{\mathrm{stack}}=\mathcal N$. The GNO magnitude $N_\alpha$ is the quotient M-circle longitudinal momentum per membrane and is identified as the dimension of the BMN block \cite{SupplementalMaterial}.

\emph{Higgs pole selection.---}
In the fixed-GNO sector specified by Eq.~\eqref{eq:abjm-saddle}, let $z_{\alpha,p}$ and $w_{\alpha,p}$ be the corresponding left and right gauge holonomies. The $U(r)_k\times U(r)_{-k}$ ABJM index \cite{Kim:2009wb} has the schematic form
\begin{align}
 \I_{\ABJM,k}^{\{n_\alpha,N_\alpha\}}
 ={}&\frac{1}{\prod_\alpha(n_\alpha!)^2}
 \oint[\dd z\,\dd w]~\mathcal Z_{\rm vec}
 \nonumber\\[-2pt]
 &\times\prod_{\alpha,\beta;p,p'}
 \Phi_{d_{\alpha\beta}}\!\left(\frac{z_{\alpha,p}}{w_{\beta,p'}}\right)
 \prod_{\alpha,p}
 \left(\frac{w_{\alpha,p}}{z_{\alpha,p}}\right)^{kN_\alpha}.
 \label{eq:abjm-index-schematic}
\end{align}
Here and below,
\begin{equation}
 d_{\alpha\beta}=|N_\alpha-N_\beta|.
 \label{eq:relative-flux}
\end{equation}
Here $\mathcal Z_{\rm vec}$ denotes the two vector-multiplet determinants, while $\Phi_d$ is the bifundamental matter one-loop factor in relative flux $d$.  The one-loop factors are independent of $k$; all explicit level dependence in this fixed sector is carried by the classical Chern--Simons monomial.  The full factors and contour are recorded in the Supplemental Material \cite{SupplementalMaterial}.  Evaluating the $w$ contours first, the relevant polar part is
\begin{equation}
 \Phi_d(u)\ \propto\
 \frac{1}{(q^{d/2}u;q)_\infty(cq^{d/2}u;q)_\infty},
 \qquad q=abc,
 \label{eq:polar-part}
\end{equation}
where $(x;q)_\infty=\prod_{n=0}^{\infty}(1-xq^n)$.  The enclosed matter poles are
\begin{equation}
 w_{\beta,p'}=\rho z_{\alpha,p},
 \qquad
 \rho=q^{d_{\alpha\beta}/2+\ell}
 \ \text{or}\ 
 cq^{d_{\alpha\beta}/2+\ell},
 \quad \ell\geq0.
 \label{eq:pole-families}
\end{equation}
At such a residue, the Chern--Simons monomial contributes
$\rho^{kN_\beta}$. In the regulated plane-wave family of Eq.~\eqref{eq:limit-shorthand}, the common boost is large, while relative membrane momenta and the corresponding light off-diagonal modes remain finite.  In the formal expansion about $a=b=c=0$, every enclosed pole other than $\rho=1$ has an exact ratio $\rho$ of positive fugacity grade.  The zero-grade ratio occurs precisely at $d_{\alpha\beta}=0$, $\ell=0$, and $\rho=1$, selecting
\begin{equation}
 w_{\alpha,p}=z_{\alpha,\sigma_\alpha(p)},
 \qquad \sigma_\alpha\in S_{n_\alpha}.
 \label{eq:w-equals-z}
\end{equation}
All permutations give the same contribution, so we choose the identity permutation, $w_{\alpha,p}=z_{\alpha,p}$, and cancel one of the two Weyl factors $\prod_\alpha n_\alpha!$ in the denominator in Eq.~\eqref{eq:abjm-index-schematic}. The pole at $w=z$ equates the left and right holonomies, as expected from the breaking to the diagonal gauge group in Eq.~\eqref{eq:classical-locking}. 

The factor $\rho^{kN_\beta}$ has a simple physical interpretation. In a sector with monopole charge $N_\beta$, the Gauss law requires the bare monopole to be dressed by $kN_\beta$ bifundamental quanta so that the dressed state is gauge invariant. A pole $w_{\beta,p'}=\rho z_{\alpha,p}$ selects the mode used for this macroscopic dressing and gives one factor of $\rho$ for every charge unit.  For $\rho=1$,
the dressing mode is the lowest $Y^4$ zero mode, and the resulting
condensate is the plane-wave saddle chosen above.  A pole with $\rho\neq1$ instead places $O(kN_\beta)$ charge units in another mode and describes a different macroscopic large-charge background rather than a perturbative fluctuation around this saddle.

Define the plane-wave contribution by summing these residues,
\begin{equation}
 \I_{\ABJM,k;\Lambda}^{\rm pw}
 \equiv\sum_{\{\sigma_\alpha\}}
 \Res_{w_{\alpha,p'}=z_{\alpha,\sigma_\alpha(p')}}
 \I_{\ABJM,k}^{\{n_\alpha,N_\alpha\}}.
 \label{eq:pw-index-definition}
\end{equation}
On these residues the Chern--Simons monomial is exactly one, and the remaining integrand is level-independent.  Consequently,
\begin{equation}
 \I_{\ABJM,k;\Lambda}^{\rm pw}
 =\I_{\ABJM,1;\Lambda}^{\rm pw}.
 \label{eq:pw-k-independence}
\end{equation}
The full remainder is controlled by expanding the one-loop integrand at fixed total grade $D$. After removing the zero-grade matter factors and within-block vector determinants, each coefficient is a finite Laurent polynomial whose exponent in any $w_{\alpha,p}$ is at least $-\lfloor D\rfloor$. The removed factors, together with the Chern--Simons monomial and measure, then make the $w_{\alpha,p}$ integrand regular at the origin whenever $kN_\alpha\geq\lfloor D\rfloor$. Its only remaining poles are those in Eq.~\eqref{eq:w-equals-z}. Thus every coefficient of $\I_{\ABJM,k;\Lambda}-\I_{\ABJM,k;\Lambda}^{\rm pw}$ at total grade less than $1+k\min_\alpha N_\alpha$ vanishes. The proof, including the remaining holonomy projection, is given in the Supplemental Material \cite{SupplementalMaterial}. Since $N_\alpha=\Lambda+\nu_\alpha$ with fixed offsets, this finite-grade bound proves coefficientwise decoupling for every fixed $k$.

After the residue sum, the comparison can be written as a single diagonal-holonomy integral.  Set $u_{\alpha,p}=(-1)^{N_\alpha}z_{\alpha,p}$ and
$\Delta(u_\alpha)=\prod_{p\neq p'}(1-u_{\alpha,p}/u_{\alpha,p'})$, and define
\begin{align}
 \I[H]={}&\frac{1}{\prod_\alpha n_\alpha!}
 \oint\prod_{\alpha,p}\frac{\dd u_{\alpha,p}}{2\pi i u_{\alpha,p}}
 \prod_\alpha\Delta(u_\alpha)
 \nonumber\\[-2pt]
 &\times\prod_{\alpha,\beta}
 \prod_{p=1}^{n_\alpha}\prod_{p'=1}^{n_\beta}
 H_{\alpha\beta}\!\left(\frac{u_{\beta,p'}}{u_{\alpha,p}}\right).
 \label{eq:common-holonomy-integral}
\end{align}
Then $\I_{\ABJM,k;\Lambda}^{\rm pw}=\I[H^{\rm mono}]$, whereas
$\I_{\BMN;\Lambda}=\I[H^{\rm mat}]$.  The kernels $H^{\rm mono}$ and
$H^{\rm mat}$ are defined below in Eqs.~\eqref{eq:monopole-kernel} and
\eqref{eq:matrix-kernel}, respectively.  The surviving ABJM kernel is therefore universal across fixed levels, and the comparison reduces to a mode-by-mode comparison of two block kernels. The detailed derivation of the reduction of the two-matrix integral of the ABJM index to the single diagonal-holonomy integral is given in the Supplemental Material \cite{SupplementalMaterial}.

\emph{Monopole harmonics versus matrix harmonics.---}
After taking the residues in Eq.~\eqref{eq:w-equals-z}, both indices are matrix integrals over the same diagonal holonomies. Their remaining distinction can be read directly from the angular harmonics connecting blocks $\alpha$ and $\beta$. With $d_{\alpha\beta}$ defined in Eq.~\eqref{eq:relative-flux}, let
\begin{equation}
 L_{\alpha\beta}=\min(N_\alpha,N_\beta).
 \label{eq:dL}
\end{equation}
Let \(\mathcal L^{d_{\alpha\beta}}\) be the monopole line bundle over \(S^2\) of degree \(d_{\alpha\beta}\), and let \(\Gamma(\mathcal L^{d_{\alpha\beta}})\) denote its space of sections. On the BMN side, let \(\Hom(\mathbb C^{N_\beta},\mathbb C^{N_\alpha})\) denote the \(N_\alpha\times N_\beta\) rectangular matrix block connecting the \(\beta\)th and \(\alpha\)th irreducible fuzzy-sphere blocks. The two harmonic spaces decompose as
\begin{align}
 \Gamma(\mathcal L^{d_{\alpha\beta}})
 &\simeq\bigoplus_{\ell=0}^{\infty}
 \mathcal R_{d_{\alpha\beta}/2+\ell},
 \nonumber\\[-2pt]
 \Hom(\mathbb C^{N_\beta},\mathbb C^{N_\alpha})
 &\simeq\bigoplus_{\ell=0}^{L_{\alpha\beta}-1}
 \mathcal R_{d_{\alpha\beta}/2+\ell},
 \label{eq:harmonic-ranges}
\end{align}
where $\mathcal R_j$ is the spin-$j$ representation of spatial $SU(2)$.  The towers have the same lower endpoint, while the fuzzy-matrix tower terminates at
$j=(N_\alpha+N_\beta)/2-1$ \cite{Ishii:2008fb,Kovacs:2013una}.  Physically, ABJM uses continuum monopole harmonics on the spherical membrane, whereas BMN uses their fuzzy-sphere matrix regularization.  The relative flux fixes the common lower spin, while finite matrix size truncates the tower only from above. Their cubic fusion algebras converge as well: the continuum monopole vertex contains two Wigner $3j$ symbols, whereas the fuzzy-matrix vertex replaces the charge $3j$ symbol by a Wigner $6j$ symbol whose large-$\Lambda$ asymptotic restores the monopole coefficient; explicit formulas are given in the Supplemental Material \cite{SupplementalMaterial}.

This statement persists for the fully refined one-loop index \cite{Chang:2024lkw,Chang:2026mui}.  Define
\begin{equation}
 \mathcal K(y)=
 \frac{(1-y)(1-aby)(1-acy)(1-bcy)}
 {(1-ay)(1-by)(1-cy)(1-qy)}.
 \label{eq:common-letter-kernel}
\end{equation}
The kernels in Eq.~\eqref{eq:common-holonomy-integral} are
\begin{align}
 H^{\rm mono}_{\alpha\beta}(x)
 &= (1-x)^{-\delta_{\alpha\beta}}
 \prod_{\ell=0}^{\infty}
 \mathcal K\!\left((-1)^{d_{\alpha\beta}}
 q^{d_{\alpha\beta}/2+\ell}x\right),
 \label{eq:monopole-kernel}\\
 H^{\rm mat}_{\alpha\beta}(x)
 &= (1-x)^{-\delta_{\alpha\beta}}
 \prod_{\ell=0}^{L_{\alpha\beta}-1}
 \mathcal K\!\left((-1)^{d_{\alpha\beta}}
 q^{d_{\alpha\beta}/2+\ell}x\right).
 \label{eq:matrix-kernel}
\end{align}
All dependence on $a,b,c$ and on the relative flux is identical mode by mode.  Consequently,
\begin{equation}
 \frac{H^{\rm mono}_{\alpha\beta}(x)}
 {H^{\rm mat}_{\alpha\beta}(x)}
 =\prod_{\ell=L_{\alpha\beta}}^{\infty}
 \mathcal K\!\left((-1)^{d_{\alpha\beta}}
 q^{d_{\alpha\beta}/2+\ell}x\right),
 \label{eq:tail-ratio}
\end{equation}
which shows that, after Higgs pole selection, the remaining difference is solely the matrix-harmonic upper cutoff.  At finite momentum, the complete discrepancy separates as in \eqref{eq:two-corrections-main}. The first term vanishes coefficientwise by the fixed-grade argument above.  The second is controlled by Eq.~\eqref{eq:tail-ratio}, starts at a fugacity grade growing with $L_{\alpha\beta}\sim\Lambda$, and is independent of $k$.  Thus the Chern--Simons level moves the non-plane-wave correction to higher grade but does not change the finite-matrix harmonic cutoff. Moreover, in the scaling of Eq.~\eqref{eq:limit-shorthand}, $d_{\alpha\beta}$ stays finite while
$L_{\alpha\beta}\to\infty$.  The tail in Eq.~\eqref{eq:tail-ratio} tends coefficientwise to one.  Using the shorthand in Eq.~\eqref{eq:limit-shorthand}, this coefficientwise selection together with Eqs.~\eqref{eq:common-holonomy-integral}, \eqref{eq:monopole-kernel}, and \eqref{eq:matrix-kernel} gives \eqref{eq:main-result}. 

\emph{ABJM, BMN, and three-dimensional SYM.---}
The correspondence above does not require an intermediate Yang--Mills description. Two complementary limits nevertheless relate its ingredients to maximally supersymmetric Yang--Mills theory on $\mathbb R\times S^2$. On the ABJM side, the novel Higgs mechanism sends both the monopole charge and the Chern--Simons level $k$ to infinity at fixed ratio, keeping the Yang--Mills coupling finite and producing $\mathcal N=8$ SYM on $\mathbb R\times S^2$ \cite{Ezhuthachan:2011dx}. This differs from the fixed-$k$ plane-wave limit used here. On the BMN side, the model expanded about $n$ copies of an $N$-dimensional fuzzy-sphere vacuum becomes $U(n)$ maximally supersymmetric Yang--Mills theory on $\mathbb R\times S^2$ as $N\to\infty$, with the noncommutativity scale removed at fixed sphere radius and Yang--Mills coupling \cite{Maldacena:2002rb}. The BMN model thus provides a supersymmetric fuzzy-sphere regularization of the curved-space SYM, and the disappearance of the matrix-harmonic cutoff is the protected-index counterpart of this continuum limit.

At $k=1$, a separate proposed relation identifies the infrared fixed point of flat-space three-dimensional $\mathcal N=8$ $U(N)$ SYM with $U(N)_1\times U(N)_{-1}$ ABJM. It has been tested by comparing $S^3$ partition functions and superconformal indices in the mirror-dual description of the SYM \cite{Kapustin:2010xq, Gang:2011xp}. This relation concerns the full infrared SCFT and the sum over its sectors. It is distinct from both the curved-space SYM above, which contains the sphere-induced mass and $\phi F$ couplings, and the fixed-GNO, selected-residue plane-wave index studied here. It remains open to identify the SYM image of the fixed monopole sector and the $w=z$ Higgs pole, and to determine whether taking the plane-wave residue and the $\Lambda\to\infty$ limit commutes with the duality and whether the BMN index emerges directly from an appropriate large-charge sector of the SYM index.

\emph{Conclusion.---}
We have established a coefficientwise correspondence between the refined
supersymmetric indices of large-monopole sectors of ABJM theory and the
BMN matrix model around general multi-fuzzy-sphere vacua, establishing a precise index relation between large-charge ABJM monopole backgrounds and multi-membrane BMN vacua.

Beyond the index itself, this construction furnishes a new duality
platform in which a nontrivial sector of a three-dimensional interacting
CFT can be compared directly with matrix quantum mechanics. One natural direction is to study the $Q$-cohomology of ABJM theory in
monopole sectors and match individual
BPS classes with those around the corresponding BMN fuzzy-sphere vacua,
including the fortuitous classes introduced in
Ref.~\cite{Chang:2024fortuity} and recently studied in ABJM
\cite{Belin:2025ABJM,Behan:2025ABJM}. The matching of the cubic harmonic algebra
further suggests extending the correspondence to protected correlation
functions and operator products. More broadly, this framework may
provide a microscopic description of interacting multi-membrane dynamics
and a controlled setting for understanding how continuum membrane
geometry emerges from matrix degrees of freedom.

\begin{acknowledgments}
\emph{Acknowledgments.---}
We thank Sarthak Duary and Xuao Zhang for helpful discussions. C.C. is supported by NSFC Grant No.~12575075. K.L. is supported by the NSFC special fund for theoretical physics No. 12447108 and the national key research and development program of China No. 2020YFA0713000. K.L. would like to thank the support of the Tsinghua University Short-term Visiting Scholarship for Doctoral Students. 
\end{acknowledgments}

\paragraph*{Data availability.}
No data were created or analyzed in this study.

\bibliography{references}

\end{document}


\title{Supplemental Material for ABJM/BMN Index Matching in the Plane-Wave Limit}

\author{Chi-Ming Chang}
\email[Contact author: ]{cmchang@tsinghua.edu.cn}
\affiliation{Yau Mathematical Sciences Center (YMSC), Tsinghua University, Beijing, China}
\affiliation{Beijing Institute of Mathematical Sciences and Applications (BIMSA), Beijing, China}
\affiliation{Peng Huanwu Center for Fundamental Theory, Hefei, Anhui, China}

\author{Kangning Liu}
\email[Contact author: ]{lkn22@mails.tsinghua.edu.cn}
\affiliation{Yau Mathematical Sciences Center (YMSC), Tsinghua University, Beijing, China}
\affiliation{Department of Mathematical Sciences, Tsinghua University, Beijing, China}
\affiliation{Jefferson Physical Laboratory, Harvard University, Cambridge, Massachusetts 02138, USA}

\date{\today}
\maketitle

This Supplemental Material presents the ABJM/BMN index conventions, charge and fugacity maps, and finite-partition BMN and general-level ABJM index formulas. It reviews the scalar-dressed ABJM half-BPS background and its Gauss-law constraints, constructs the fixed-GNO index with its contour prescription, and evaluates the Higgs residues that yield the plane-wave contribution. It also compares the fusion coefficients of continuum monopole and finite matrix harmonics and provides an explicit two-block index check.

\section{Index conventions}
\label{sec:conventions}

We follow the index conventions of Refs.~\cite{Kim:2009wb,Chang:2024lkw}.
On the BMN side, let $H$ be the Hamiltonian and
$M^{12},M^{45},M^{67},M^{89}$ the commuting rotation generators.
Choose $Q=Q^4_{-}$ and set
\begin{equation}
 2\Delta_{\BMN}\equiv\{Q,Q^\dagger\}
 =2H-\frac{2\mu}{3}M^{12}
 -\frac{\mu}{3}\left(M^{45}+M^{67}+M^{89}\right).
 \label{eq:S-BMN-BPS-generator}
\end{equation}
For ABJM theory at level $k$, let $D$ be the cylinder energy, $J_{\rm rot}$
the spatial spin, and
$\mathbf M^{12},\mathbf M^{34},\mathbf M^{56},\mathbf M^{78}$ the four
commuting charges used in the index.  The subscript on
$J_{\rm rot}$ distinguishes it from the covering-space charge $J$ in
Eq.~\eqref{eq:S-covering-charge-integral}.
Here $\mathbf M^{12},\mathbf M^{34},\mathbf M^{78}$ are the Cartans of
$SO(6)_R$, while $\mathbf M^{56}$ is the monopole charge.
The Poincar\'e supercharges $\mathbf Q^7_\alpha,\mathbf Q^8_\alpha$ and
conformal supercharges $\mathbf S^{7,\alpha},\mathbf S^{8,\alpha}$ are
neutral under $\mathbf M^{56}$ and belong to the $OSp(6|4)$ subalgebra
present for every $k$.  Their anticommutator defines the BPS grading
\begin{equation}
 \Delta_{\ABJM}\equiv\frac12\left\{
 \mathbf Q^7_-+i\mathbf Q^8_-,
 \mathbf S^{7,-}-i\mathbf S^{8,-}
 \right\}
 =D-J_{\rm rot}-\mathbf M^{78}.
 \label{eq:S-ABJM-BPS-generator}
\end{equation}

At $k=1$, the BMN $SU(4|2)$ superalgebra embeds into the enhanced
$OSp(8|4)$ superconformal algebra of ABJM theory
\cite{Chang:2024lkw,Dasgupta:2002hx}.  With $A=1,\ldots,8$ labeling the
ABJM supercharges, the embedding gives
\begin{equation*}
 Q^I_\alpha=\sqrt{\frac{\mu}{3}}
 \left(\mathbf Q^{2I-1}_\alpha+i\mathbf Q^{2I}_\alpha\right),
 \qquad
 \left(Q^I_\alpha\right)^\dagger=\sqrt{\frac{\mu}{3}}
 \left(\mathbf S^{2I-1,\alpha}-i\mathbf S^{2I,\alpha}\right).
\end{equation*}
In particular, the supercharge in Eq.~\eqref{eq:S-BMN-BPS-generator}
corresponds to
\begin{equation*}
 Q^4_- =\sqrt{\frac{\mu}{3}}
 \left(\mathbf Q^7_-+i\mathbf Q^8_-\right),
 \qquad
 \left(Q^4_-\right)^\dagger=\sqrt{\frac{\mu}{3}}
 \left(\mathbf S^{7,-}-i\mathbf S^{8,-}\right).
\end{equation*}
The corresponding BMN and ABJM charges obey
\begin{equation*}
 \begin{aligned}
 \frac{3H}{\mu}
 &=D-\frac14\left(
 \mathbf M^{12}+\mathbf M^{34}+\mathbf M^{56}+\mathbf M^{78}
 \right),
 &\qquad M^{12}&=J_{\rm rot},\\
 M^{45}
 &=\frac12\left(
 \mathbf M^{12}-\mathbf M^{34}-\mathbf M^{56}+\mathbf M^{78}
 \right),
 &
 M^{67}
 &=\frac12\left(
 -\mathbf M^{12}+\mathbf M^{34}-\mathbf M^{56}+\mathbf M^{78}
 \right),\\
 M^{89}
 &=\frac12\left(
 -\mathbf M^{12}-\mathbf M^{34}+\mathbf M^{56}+\mathbf M^{78}
 \right).
 \end{aligned}
\end{equation*}
Substitution into Eqs.~\eqref{eq:S-BMN-BPS-generator} and
\eqref{eq:S-ABJM-BPS-generator} gives
$\Delta_{\BMN}=\frac{\mu}{3}\Delta_{\ABJM}$.  At $k=1$, these relations
follow from the $SU(4|2)\subset OSp(8|4)$ embedding above.  For general
$k$, they specify the correspondence between the mutually commuting
charges that occur in the two traces; no $OSp(8|4)$ superalgebra embedding
is assumed.

In the equal left--right flux sectors used in the Letter, with
$m_{\alpha,p}=\widetilde m_{\alpha,p}=-N_\alpha$, the flux convention of
Ref.~\cite{Chang:2024lkw} gives the ABJM charge that carries the M-theory
circle momentum:
\begin{equation*}
 \begin{aligned}
 \mathbf M^{56}
 &=-\frac{k}{4\pi}\int_{S^2}\Tr F
 =-\frac{k}{4\pi}\int_{S^2}\Tr\widehat F
 =-\frac{k}{2}\sum_{I=1}^r m_I
 =-\frac{k}{2}\sum_{I=1}^r\widetilde m_I,\\
 \mathbf M^{56}
 &=\frac{k}{2}\sum_\alpha n_\alpha N_\alpha
 =\frac{J}{2},
 \qquad
 \frac{2\mathbf M^{56}}{k}=\sum_\alpha n_\alpha N_\alpha.
 \end{aligned}
\end{equation*}
Thus $2\mathbf M^{56}/k$ is the integer quotient M-circle momentum
used in the Letter \cite{Kim:2009wb,Chang:2024lkw}.

With the charges fixed, let $\Delta_1,\Delta_2,\Delta_3$ be the three
independent chemical potentials conjugate to $M^{12}+M^{45}$,
$M^{12}+M^{67}$, and $M^{12}+M^{89}$, respectively.  We define the common
fugacities
\begin{equation}
 a=e^{-\Delta_1},\qquad b=e^{-\Delta_2},\qquad c=e^{-\Delta_3},
 \qquad q=abc.
 \label{eq:S-fugacities}
\end{equation}
For a BMN vacuum $\mathfrak p$, the trace used in the Letter is
\begin{align}
 \I_{\BMN}^{\mathfrak p}(a,b,c)
 =\Tr_{\mathcal H_{\mathfrak p}}\!\left[
 (-1)^{2M^{12}}e^{-\beta\Delta_{\BMN}}
 a^{M^{12}+M^{45}}
 b^{M^{12}+M^{67}}
 c^{M^{12}+M^{89}}
 \right].
 \label{eq:S-BMN-trace}
\end{align}

With chemical potentials $\gamma_1,\ldots,\gamma_4$, the ABJM sector
trace is
\begin{align}
 \I_{\ABJM,k}^{\vec m,\widetilde{\vec m}}
 &=\Tr_{\mathcal H_{\vec m,\widetilde{\vec m}}}\!\left[
 (-1)^F e^{-\beta\Delta_{\ABJM}}
 e^{-\gamma_1\mathbf M^{12}-\gamma_2\mathbf M^{34}
 -\gamma_3\mathbf M^{56}-\gamma_4(2J_{\rm rot}+\mathbf M^{78})}
 \right].
 \label{eq:S-ABJM-trace}
\end{align}

Substituting the charge relations above into Eq.~\eqref{eq:S-BMN-trace}
and comparing its charge weights with those in
Eq.~\eqref{eq:S-ABJM-trace} gives
\begin{equation}
 \begin{aligned}
 \gamma_1&=\frac{\Delta_1-\Delta_2-\Delta_3}{2},
 &\qquad
 \gamma_2&=\frac{-\Delta_1+\Delta_2-\Delta_3}{2},\\
 \gamma_3&=\frac{-\Delta_1-\Delta_2+\Delta_3}{2},
 &
 \gamma_4&=\frac{\Delta_1+\Delta_2+\Delta_3}{2}.
 \end{aligned}
 \label{eq:S-chemical-potential-map}
\end{equation}

The energy fugacity $\xi$ and flavor fugacities $y_1,y_2,y_3$ in the
standard BMN and ABJM formulas are
\begin{equation}
 \begin{aligned}
 \xi=e^{-\gamma_4}&=q^{1/2},
 &\qquad
 y_1=e^{-\gamma_1}&=\sqrt{\frac{a}{bc}},\\
 y_2=e^{-\gamma_2}&=\sqrt{\frac{b}{ac}},
 &
 y_3=e^{-\gamma_3}&=\sqrt{\frac{c}{ab}}.
 \end{aligned}
 \label{eq:S-fugacity-map}
\end{equation}
These identifications express both indices in the common variables
$(q,a,b,c)$ \cite{Chang:2024lkw}. 
We work in the chamber $|a|,|b|,|c|,|q|<1$, or equivalently
$\Delta_{1,2,3}>0$.

For a complex variable $x$ and a nonnegative integer $n$, we use
\begin{equation}
 (x;q)_n=\prod_{\ell=0}^{n-1}(1-xq^\ell),
 \qquad
 (x;q)_\infty=\prod_{\ell=0}^{\infty}(1-xq^\ell),
 \label{eq:S-pochhammer}
\end{equation}
and
\begin{equation}
 \PE[f(x_1,\ldots,x_t)]
 =\exp\left[\sum_{m=1}^{\infty}\frac1m
 f(x_1^m,\ldots,x_t^m)\right].
 \label{eq:S-PE}
\end{equation}
We repeatedly use
\begin{equation}
 (x;q)_\infty=(1-x)(qx;q)_\infty,
 \quad
 \PE\left[\frac{x}{1-q}\right]=\frac1{(x;q)_\infty},
 \quad
 \PE\left[-\frac{x}{1-q}\right]=(x;q)_\infty.
 \label{eq:S-basic-identities}
\end{equation}
\subsection{Finite index formulas}
\label{sec:S-index-formulas}

For two partition blocks labeled by $\alpha$ and $\beta$, define
\begin{equation}
 d_{\alpha\beta}=|N_\alpha-N_\beta|,
 \qquad
 L_{\alpha\beta}=\min(N_\alpha,N_\beta).
 \label{eq:S-dL}
\end{equation}
Let $u_{\alpha,p}$, $p=1,\ldots,n_\alpha$, be the eigenvalue holonomies of
the unbroken $U(n_\alpha)$ factor, integrated counterclockwise on the unit
circle.  The BMN index for the partition data specified in the Letter is
\cite{Chang:2024lkw,Chang:2026mui}
\begin{align}
 \I_{\BMN}^{\{n_\alpha,N_\alpha\}}
 ={}&\frac{1}{\prod_\alpha n_\alpha!}
 \oint\prod_{\alpha,p}\frac{\dd u_{\alpha,p}}{2\pi i u_{\alpha,p}}
 \prod_\alpha\Delta(u_\alpha)
 \nonumber\\[-2pt]
 &\times\exp\left[
 \sum_{m=1}^{\infty}\frac1m
 \sum_{\alpha,\beta}\widehat\iota_{\alpha\beta}^{(m)}
 \sum_{p=1}^{n_\alpha}\sum_{p'=1}^{n_\beta}
 \left(\frac{u_{\beta,p'}}{u_{\alpha,p}}\right)^m
 \right],
 \label{eq:S-bmn-index-exp}
\end{align}
where
\begin{equation}
 \Delta(u_\alpha)=\prod_{p\neq p'}
 \left(1-\frac{u_{\alpha,p}}{u_{\alpha,p'}}\right),
 \label{eq:S-haar}
\end{equation}
and
\begin{equation}
 \widehat\iota_{\alpha\beta}^{(m)}
 =\delta_{\alpha\beta}
 -(1-a^m)(1-b^m)(1-c^m)
 (-1)^{m d_{\alpha\beta}}q^{m d_{\alpha\beta}/2}
 \frac{1-q^{mL_{\alpha\beta}}}{1-q^m}.
 \label{eq:S-letter-m}
\end{equation}
Here $\delta_{\alpha\beta}$ is the Kronecker delta.

For $U(r)_k\times U(r)_{-k}$ ABJM theory, let
$m_I,\widetilde m_I\in\mathbb Z$, $I=1,\ldots,r$, denote the left and
right GNO charges.  The ABJM index is a sum over GNO sectors and an
integral over the corresponding gauge holonomies
\cite{Kim:2009wb,Chang:2024lkw}.  In the angular variables
$\alpha_I,\widetilde\alpha_I\in[-\pi,\pi]$ used in
\cite{Chang:2024lkw}, the multiplicative holonomies are
\begin{equation}
 Z_I=e^{i\alpha_I},
 \qquad
 W_I=e^{i\widetilde\alpha_I}.
 \label{eq:S-holonomy-angular-map}
\end{equation}
We distinguish the holonomy variables $Z_I$ from the bifundamental scalar
$Z$ used in the half-BPS derivation below.  Equivalently,
\begin{equation}
 \frac{\dd\alpha_I}{2\pi}
 =\frac{\dd Z_I}{2\pi i Z_I},
 \qquad
 \frac{\dd\widetilde\alpha_I}{2\pi}
 =\frac{\dd W_I}{2\pi i W_I},
 \qquad
 e^{ik(m_I\alpha_I-\widetilde m_I\widetilde\alpha_I)}
 =Z_I^{km_I}W_I^{-k\widetilde m_I}.
 \label{eq:S-holonomy-measure-map}
\end{equation}
The full ABJM index is
\begin{align}
 \I_{\ABJM,k}^{(r)}
 ={}&
 \frac{1}{(r!)^2}
 \sum_{m_I,\widetilde m_I\in\mathbb Z}
 y_3^{-\frac{k}{2}\sum_{I=1}^r m_I}
 \oint
 \prod_{I=1}^r
 \frac{\dd Z_I}{2\pi i Z_I}
 \frac{\dd W_I}{2\pi i W_I}
 \prod_{I=1}^r Z_I^{km_I}W_I^{-k\widetilde m_I}
 \nonumber\\
 &\times
 \xi^{\epsilon_0}
 \exp\left[
 \sum_{n=1}^{\infty}\frac{1}{n}
 f\!\left(
 \xi^n,y_1^n,y_2^n,
 \{Z_I^n\},\{W_I^n\}
 \right)
 \right].
 \label{eq:S-full-abjm-index}
\end{align}
The fugacities in this formula were defined in
Eq.~\eqref{eq:S-fugacity-map}.  The zero-point energy is
\begin{equation}
 \epsilon_0
 =
 \sum_{I,J}|m_I-\widetilde m_J|
 -\sum_{I<J}|m_I-m_J|
 -\sum_{I<J}|\widetilde m_I-\widetilde m_J|.
 \label{eq:S-epsilon0}
\end{equation}
The single-letter expression is
\begin{align}
 f={}&
 -\sum_{I\neq J}
 \xi^{|m_I-m_J|}\frac{Z_J}{Z_I}
 -\sum_{I\neq J}
 \xi^{|\widetilde m_I-\widetilde m_J|}
 \frac{W_J}{W_I}
 \nonumber\\
 &+\sum_{I,J}
 \xi^{|m_I-\widetilde m_J|}
 \left(
 f_+\frac{W_J}{Z_I}
 +f_-\frac{Z_I}{W_J}
 \right),
 \label{eq:S-abjm-letter}
\end{align}
where
\begin{align}
 f_+(\xi,y_1,y_2)
 &=
 \frac{1}{1-\xi^2}
 \left[
 \xi^{1/2}
 \left(
 \sqrt{\frac{y_1}{y_2}}
 +\sqrt{\frac{y_2}{y_1}}
 \right)
 \right.
 \nonumber\\[-2pt]
 &\hspace{27mm}\left.
 -\xi^{3/2}
 \left(
 \sqrt{y_1y_2}
 +\frac{1}{\sqrt{y_1y_2}}
 \right)
 \right],
 \\
 f_-(\xi,y_1,y_2)
 &=f_+(\xi,y_1,y_2^{-1}).
 \label{eq:S-fpm}
\end{align}

\section{Derivation of the ABJM half-BPS background}
\label{sec:S-bps-backgrounds}

The Letter states the BMN/ABJM vacuum and momentum dictionary.  Here we review the derivation of the ABJM half-BPS solutions following \cite{SheikhJabbari:2009ns,Ezhuthachan:2011dx}.  The theory has gauge group $U(r)_k\times U(r)_{-k}$, Hermitian gauge connections $A_\mu$ and $\widehat A_\mu$, and four complex bifundamental scalars $Y^A$.  We use
\begin{equation}
 D_\mu Y^A=\partial_\mu Y^A+iA_\mu Y^A-iY^A\widehat A_\mu,
 \qquad
 F=\dd A+iA\wedge A,
 \qquad
 \widehat F=\dd\widehat A+i\widehat A\wedge\widehat A.
 \label{eq:S-abjm-conventions}
\end{equation}
Repeated $SU(4)$ R-symmetry indices $A,B,C=1,\ldots,4$ are summed.  The potential $V_6(Y)$ is positive-semidefinite as it can be rewritten as a sum
of squares; it also vanishes when all $Y^A$
are proportional to a single commuting matrix. On $\mathbb R_t\times S^2_L$, the bosonic action is
\begin{align}
 S_{\rm bos}={}&\frac{k}{4\pi}\int
 \Tr\left(A\wedge\dd A+\frac{2i}{3}A\wedge A\wedge A
 -\widehat A\wedge\dd\widehat A
 -\frac{2i}{3}\widehat A\wedge\widehat A\wedge\widehat A\right)
 \nonumber\\
 &-\int\dd^3x\sqrt{-g}\,\Tr\left(
 D_\mu Y_A^\dagger D^\mu Y^A+\frac{1}{4L^2}Y_A^\dagger Y^A\right)
 -\int\dd^3x\sqrt{-g}\,V_6(Y),
 \label{eq:S-abjm-action}
\end{align}
where the conformal mass is fixed by the sphere radius and $V_6$ is the positive-semidefinite sextic scalar potential \cite{Aharony:2008ug}.  The spherically symmetric ABJM saddle used in the Letter lies on the half-BPS moduli space derived below.

A point on the half-BPS moduli space is specified by a unit vector
$u^A$ in the scalar R-symmetry space, normalized by
$\bar u_Au^A=1$.  Define $Z\equiv\bar u_A Y^A$ and
$Y_\perp^A\equiv(\delta^A{}_B-u^A\bar u_B)Y^B$.  The BPS Hamiltonian is understood to
use the $U(1)$ charge that rotates this complex direction.  
The BPS bound relevant for this chiral branch can be made explicit as
follows.  Restricting first to bosonic configurations, the Chern--Simons
gauge fields carry no local Hamiltonian density, and the scalar contribution
to the cylinder Hamiltonian is
\begin{equation}
 H_{\rm scal}
 =
 \int_{S^2}\dd^2x\,\sqrt g\,\left\{
 \Tr\left[
 D_0Y_A^\dagger D_0Y^A
 +D_iY_A^\dagger D^iY^A
 +\frac{1}{4L^2}Y_A^\dagger Y^A
 \right]+V_6(Y)
 \right\}.
 \label{eq:S-scalar-Hamiltonian}
\end{equation}
Here $g$ denotes the determinant of the metric on $S^2_L$.  The term
$1/(4L^2)$ is the conformal mass of a scalar on
$\mathbb R_t\times S^2_L$.  The temporal gauge fields $A_0$ and
$\widehat A_0$ instead act as Lagrange multipliers and impose the
Chern--Simons Gauss constraints, which will be discussed below.

For a chosen unit vector $u^A$, the $U(1)\subset SU(4)_R$ that defines
the chiral direction is generated by the traceless matrix
\begin{equation}
 (T_u)^A{}_B
 =
 \frac{4}{3}u^A\bar u_B-\frac13\delta^A{}_B.
 \label{eq:S-Tu-generator}
\end{equation}
With
\begin{equation}
 Y^A=u^A Z+Y_\perp^A,
 \qquad
 Z=\bar u_A Y^A,
 \qquad
 \bar u_A Y_\perp^A=0,
 \label{eq:S-u-decomposition}
\end{equation}
its action is
\begin{equation}
 T_u (uZ)=uZ,
 \qquad
 T_u(Y_\perp)=-\frac13Y_\perp.
 \label{eq:S-Tu-charges}
\end{equation}
Thus our normalization assigns $U(1)_u$ charge $+1$ to the scalar
along the chiral direction and charge $-1/3$ to each of the three
orthogonal complex scalars.

For any complex scalar matrix $\Phi$, define its phase charge by
\begin{equation}
 j[\Phi]
 \equiv
 -i\int_{S^2}\dd^2x\,\sqrt g\,
 \Tr\left[
 (D_0\Phi)\Phi^\dagger
 -\Phi(D_0\Phi^\dagger)
 \right].
 \label{eq:S-phase-charge}
\end{equation}
The Noether charge associated with $T_u$ is therefore
\begin{equation}
 J_u
 =
 j[Z]-\frac13 j[Y_\perp],
 \label{eq:S-Ju}
\end{equation}
where
\begin{equation}
 j[Y_\perp]
 \equiv
 -i\int_{S^2}\dd^2x\,\sqrt g\,
 \Tr\left[
 (D_0Y_\perp^A)Y_{\perp A}^\dagger
 -Y_\perp^A(D_0Y_{\perp A}^\dagger)
 \right],
 \label{eq:S-Jperp}
\end{equation}
with the orthogonal $SU(4)$ index summed.

For the spinless chiral sector preserving $SO(3)$, the superconformal BPS bound is
\begin{equation}
 H_{\rm scal}\geq \frac{1}{2L}J_u.
 \label{eq:S-BPS-bound}
\end{equation}
On the bosonic scalar configurations considered here, it is therefore
convenient to define
\begin{equation}
 H_{\rm BPS}
 \equiv
 H_{\rm scal}-\frac{1}{2L}J_u.
 \label{eq:S-BPS-H-def}
\end{equation}
The coefficient $1/(2L)$ is fixed by the cylinder energy of a chiral
scalar: a scalar of $U(1)_u$ charge one has conformal energy
$1/(2L)$.

Using the decomposition in Eq.~\eqref{eq:S-u-decomposition}, the
quadratic part of the scalar Hamiltonian separates into the $Z$ and
$Y_\perp$ sectors.  For the chiral scalar $Z$ one finds
\begin{align}
 &\Tr\left[
 D_0Z^\dagger D_0Z
 +\frac{1}{4L^2}Z^\dagger Z
 \right]
 -\frac{1}{2L}\,
 \left\{
 -i\Tr\left[
 (D_0Z)Z^\dagger
 -Z(D_0Z^\dagger)
 \right]
 \right\}
 \nonumber\\
 &\hspace{15mm}
 =
 \Tr\left[
 \left(D_0Z-\frac{i}{2L}Z\right)^\dagger
 \left(D_0Z-\frac{i}{2L}Z\right)
 \right].
 \label{eq:S-Z-square}
\end{align}
The conformal mass is therefore exactly absorbed into the time-derivative
square in the charge-one direction.

For the orthogonal scalars, whose $U(1)_u$ charge is $-1/3$, the
corresponding combination is instead
\begin{align}
 &\Tr\left[
 D_0Y_{\perp A}^\dagger D_0Y_\perp^A
 +\frac{1}{4L^2}Y_{\perp A}^\dagger Y_\perp^A
 \right]
 +\frac{1}{6L}\,
 \left\{
 -i\Tr\left[
 (D_0Y_\perp^A)Y_{\perp A}^\dagger
 -Y_\perp^A(D_0Y_{\perp A}^\dagger)
 \right]
 \right\}
 \nonumber\\
 &\hspace{8mm}
 =
 \Tr\left[
 \left(D_0Y_\perp^A+\frac{i}{6L}Y_\perp^A\right)^\dagger
 \left(D_0Y_\perp^A+\frac{i}{6L}Y_\perp^A\right)
 +\frac{2}{9L^2}Y_{\perp A}^\dagger Y_\perp^A
 \right].
 \label{eq:S-Yperp-square}
\end{align}
The last term is strictly positive for nonzero $Y_\perp^A$.  Combining
Eqs.~\eqref{eq:S-Z-square} and \eqref{eq:S-Yperp-square}, together with
the spatial gradients and the positive-semidefinite sextic potential,
gives
\begin{align}
 H_{\rm BPS}
 =
 \int_{S^2}\dd^2x\,\sqrt g\,\Bigg\{\Tr\Bigg[
 &
 \left(D_0Z-\frac{i}{2L}Z\right)^\dagger
 \left(D_0Z-\frac{i}{2L}Z\right)
 +D_iZ^\dagger D^iZ
 \nonumber\\
 &+
 \left(D_0Y_\perp^A+\frac{i}{6L}Y_\perp^A\right)^\dagger
 \left(D_0Y_\perp^A+\frac{i}{6L}Y_\perp^A\right)
 +D_iY_{\perp A}^\dagger D^iY_\perp^A
 \nonumber\\
 &+
 \frac{2}{9L^2}Y_{\perp A}^\dagger Y_\perp^A
 \Bigg]+V_6(Y)\Bigg\}.
 \label{eq:S-BPS-Hamiltonian}
\end{align}
Every term on the right-hand side is nonnegative.  Saturation of the
BPS bound, $H_{\rm BPS}=0$, therefore requires
\begin{equation}
 D_iZ=0,
 \qquad
 D_0Z=\frac{i}{2L}Z,
 \qquad
 Y_\perp^A=0,
 \qquad
 V_6(Y)=0.
 \label{eq:S-cylinder-BPS-equations}
\end{equation}
Here and below
$i=1,2$ labels directions on $S^2$. In addition to the algebraic condition $V_6=0$, the spatial equation constrains how the scalar zero mode can connect magnetic bundles.  Acting with the R-symmetry changes $u^A$ and the associated BPS generator but leaves the orbit of solutions unchanged.

Define the two GNO matrices by
\begin{equation}
 M=\frac{1}{2\pi}\int_{S^2}F,
 \qquad
 \widehat M=\frac{1}{2\pi}\int_{S^2}\widehat F.
 \label{eq:S-GNO-matrices}
\end{equation}
To spell out the integrability step, commuting two spatial covariant
derivatives gives
\begin{equation*}
 0=[D_i,D_j]Z=i\big(F_{ij}Z-Z\widehat F_{ij}\big).
\end{equation*}
Thus the covariantly constant bundle map $Z$ intertwines the two curvature
operators.  For diagonal GNO representatives in the standard polar
coordinates $(\theta,\phi)$ on $S^2$,
$F=\tfrac12M\sin\theta\,\dd\theta\wedge\dd\phi$ and
$\widehat F=\tfrac12\widehat M\sin\theta\,\dd\theta\wedge\dd\phi$,
so the preceding pointwise identity gives $MZ=Z\widehat M$.  Writing
$M=\diag(m_1,\ldots,m_r)$ and
$\widehat M=\diag(\widehat m_1,\ldots,\widehat m_r)$ then gives, component
by component,
\begin{equation}
 MZ=Z\widehat M,
 \qquad
 (m_i-\widehat m_j)Z_{ij}=0.
 \label{eq:S-flux-intertwiner}
\end{equation}
where $Z_{ij}$ maps the $j$th right GNO eigenspace to the $i$th left one.
Thus $Z_{ij}\neq0$ requires equal left and right flux eigenvalues, $m_i=\widehat m_j$.

The matter charge densities are
\begin{align}
 \rho_L&=i\sum_A\left[(D_0Y^A)Y_A^\dagger-Y^A(D_0Y_A^\dagger)\right],
 \nonumber\\
 \rho_R&=i\sum_A\left[(D_0Y_A^\dagger)Y^A-Y_A^\dagger(D_0Y^A)\right].
 \label{eq:S-rho-def}
\end{align}
Variation of $A_0$ and $\widehat A_0$ gives the two matrix-valued Gauss laws
\begin{equation}
{
 \frac{k}{2\pi}F=\rho_L\,\vol_{S^2},
 \qquad
 \frac{k}{2\pi}\widehat F=-\rho_R\,\vol_{S^2}
 }.
 \label{eq:S-matrix-gauss}
\end{equation}
On the half-BPS branch in Eq.~\eqref{eq:S-cylinder-BPS-equations}, these reduce to the two moment-map equations
\begin{equation}
 kM=-4\pi L\,ZZ^\dagger,
 \qquad
 k\widehat M=-4\pi L\,Z^\dagger Z.
 \label{eq:S-two-moment-maps}
\end{equation}
The nonzero spectra of $ZZ^\dagger$ and $Z^\dagger Z$ agree, so the left and right multiplicities of every nonzero flux agree. Since the two gauge groups have equal rank, the zero-flux multiplicities agree as well. Hence the GNO spectra coincide up to permutation, and a right Weyl permutation can be used to set $M=\widehat M$.

For the fixed-GNO sector associated with the block data $(n_\alpha, N_\alpha)$ in the Letter, take
\begin{equation}
 r=\sum_{\alpha=1}^{s}n_\alpha,
 \qquad
 M=\widehat M=-\bigoplus_{\alpha=1}^{s}
 N_\alpha\mathbf 1_{n_\alpha},
 \label{eq:S-negative-flux}
\end{equation}
where the sign is the convention used in the index calculation and consistent with \eqref{eq:S-two-moment-maps}. A northern-patch representative is
\begin{equation}
 A^{(0)}=\widehat A^{(0)}
 =-\frac12\left(\bigoplus_\alpha N_\alpha\mathbf 1_{n_\alpha}\right)
 (1-\cos\theta)\,\dd\phi.
 \label{eq:S-monopole-connection}
\end{equation}
A scalar dressing is required for the flux to satisfy Eq.~\eqref{eq:S-matrix-gauss}.  For arbitrary
unit $u^A$, the most general half-BPS dressing compatible with the
unsplit multiplicities $n_\alpha$ is parametrized by complex amplitudes
$v_\alpha$ and unitary relative orientations $U_\alpha$:
\begin{equation}
 Y^A_{(0)}(t)=u^A e^{+it/(2L)}
 \bigoplus_{\alpha=1}^{s}v_\alpha U_\alpha,
 \qquad
 U_\alpha\in U(n_\alpha),
 \qquad
 A_0^{(0)}=\widehat A_0^{(0)}=0.
 \label{eq:S-scalar-dressing}
\end{equation}
The unitary matrices $U_\alpha$ are removed by blockwise left--right gauge transformations.  We henceforth set $U_\alpha=\mathbf 1_{n_\alpha}$ and use a global $SU(4)$ rotation to choose $u^A=\delta^{A4}$, so that $Y^4_{(0)}\neq0$ and $Y^{A\neq4}_{(0)}=0$. The Gauss law \eqref{eq:S-two-moment-maps} fixes, block by block,
\begin{equation}
{|v_\alpha|^2=\frac{kN_\alpha}{4\pi L}}.
 \label{eq:S-scalar-amplitude}
\end{equation}
This is a solution of the complete classical equations. 

The bare monopole flux preserves
\begin{equation}
 \prod_{\alpha}U(n_\alpha)_L\times U(n_\alpha)_R.
 \label{eq:S-flux-centralizer}
\end{equation}
The scalar condensate is invariant only when
\begin{equation}
 g_{L,\alpha}Y^4_{(0)}g_{R,\alpha}^{-1}=Y^4_{(0)},
 \label{eq:S-condensate-invariance}
\end{equation}
which, in the gauge chosen above, requires $g_{L,\alpha}=g_{R,\alpha}$.  It therefore breaks the two gauge factors to the diagonal subgroup
\begin{equation}
\prod_\alpha\left[U(n_\alpha)_L\times U(n_\alpha)_R\right]
 \longrightarrow \prod_\alpha U(n_\alpha)_{\rm diag}.
 \label{eq:S-broken-gauge-group}
\end{equation}
On the chosen representative, the covering-space scalar charge is realized as
\begin{align}
 J
 &\equiv-i\int_{S^2}\dd^2x\sqrt g\,
 \Tr\left[(D_0Y^4_{(0)})(Y^4_{(0)})^\dagger
 -Y^4_{(0)}D_0(Y^4_{(0)})^\dagger\right]
 \nonumber\\[-2pt]
 &=\sum_{\alpha=1}^{s}4\pi L n_\alpha|v_\alpha|^2.
 \label{eq:S-covering-charge-integral}
\end{align}
The contribution of the $\alpha$th stack and the corresponding quotient M-circle momentum are therefore
\begin{equation}
 J_\alpha^{\mathrm{stack}}=k n_\alpha N_\alpha,
 \qquad
 P_\alpha^{\mathrm{stack}}=\frac{J_\alpha^{\mathrm{stack}}}{k}=n_\alpha N_\alpha.
 \label{eq:S-block-momenta}
\end{equation}
This establishes the charge
and momentum dictionary stated in the Letter. In the remainder we use the Letter’s
fixed-difference large-momentum family and hold the Chern–Simons level fixed. 

\section{Finite-sector kernels and fixed-GNO ABJM reduction}
\label{sec:S-finite-indices}

For the BMN index in Eq.~\eqref{eq:S-bmn-index-exp}, define
\begin{equation}
 \begin{gathered}
 \mathcal K(y)=
 \frac{(1-y)(1-aby)(1-acy)(1-bcy)}
 {(1-ay)(1-by)(1-cy)(1-qy)},\\
 H^{\rm mat}_{\alpha\beta}(x)
 =(1-x)^{-\delta_{\alpha\beta}}
 \prod_{\ell=0}^{L_{\alpha\beta}-1}
 \mathcal K\!\left((-1)^{d_{\alpha\beta}}
 q^{d_{\alpha\beta}/2+\ell}x\right).
 \end{gathered}
 \label{eq:S-matrix-kernel}
\end{equation}
Introduce the common diagonal-holonomy integral
\begin{align}
 \I[H]={}&\frac{1}{\prod_\alpha n_\alpha!}
 \oint\prod_{\alpha,p}\frac{\dd u_{\alpha,p}}{2\pi i u_{\alpha,p}}
 \prod_\alpha\Delta(u_\alpha)
 \nonumber\\[-2pt]
 &\times\prod_{\alpha,\beta}
 \prod_{p=1}^{n_\alpha}\prod_{p'=1}^{n_\beta}
 H_{\alpha\beta}\!\left(\frac{u_{\beta,p'}}{u_{\alpha,p}}\right).
 \label{eq:S-common-integral}
\end{align}
Then the BMN index is
\begin{equation}
 \I_{\BMN}^{\{n_\alpha,N_\alpha\}}=\I[H^{\rm mat}].
 \label{eq:S-bmn-index}
\end{equation}
For $\alpha=\beta$ and $p=p'$, Eq.~\eqref{eq:S-common-integral} sets
$x=1$.  The factors at $x=1$ are evaluated together:
\begin{align}
 H^{\rm mat}_{\alpha\alpha}(1)
 &=\lim_{x\to1}\frac{\mathcal K(x)}{1-x}
 \prod_{\ell=1}^{N_\alpha-1}\mathcal K(q^\ell x)
 \nonumber\\[-2pt]
 &=\frac{(1-ab)(1-ac)(1-bc)}
 {(1-a)(1-b)(1-c)(1-q)}
 \prod_{\ell=1}^{N_\alpha-1}\mathcal K(q^\ell).
 \label{eq:S-matrix-kernel-diagonal}
\end{align}
The product terminates because an $N_\alpha\times N_\beta$ rectangular matrix contains only finitely many angular harmonics.

For the ABJM index in Eq.~\eqref{eq:S-full-abjm-index}, we are interested
in the fixed-GNO sector
\begin{equation}
 m_{\alpha,p}=\widetilde m_{\alpha,p}=-N_\alpha,
 \qquad
 p=1,\ldots,n_\alpha.
 \label{eq:S-fixed-gno}
\end{equation}
This implies$\epsilon_0=0$. The residual Weyl factor is $1/\prod_\alpha(n_\alpha!)^2$. Rescale
\begin{equation}
 Z_{\alpha,p}=\sqrt{y_3}\,z_{\alpha,p},
 \qquad W_{\alpha,p}=w_{\alpha,p}.
 \label{eq:S-holonomy-rescale}
\end{equation}
Substituting Eqs.~\eqref{eq:S-fixed-gno} and
\eqref{eq:S-holonomy-rescale} into the $y_3$ and holonomy powers in
Eq.~\eqref{eq:S-full-abjm-index} gives
\begin{align}
 &y_3^{\frac{k}{2}\sum_\alpha n_\alpha N_\alpha}
 \prod_{\alpha,p}
 (\sqrt{y_3}\,z_{\alpha,p})^{-kN_\alpha}
 w_{\alpha,p}^{kN_\alpha}
 \nonumber\\[-2pt]
 &\qquad=\prod_{\alpha,p}
 z_{\alpha,p}^{-kN_\alpha}w_{\alpha,p}^{kN_\alpha}.
 \label{eq:S-CS-phase}
\end{align}
Moreover, Eq.~\eqref{eq:S-fugacity-map} gives
\begin{equation}
 y_3^{-1/2}f_+=\frac{a+b-ab-q}{1-q},
 \qquad
 y_3^{1/2}f_-=\frac{1+c-ac-bc}{1-q}.
 \label{eq:S-rescaled-letters}
\end{equation}
For $d\geq0$, define
\begin{align}
 \Phi_d(u)
 &\equiv\PE\left[q^{d/2}\left(
 \frac{a+b-ab-q}{1-q}u^{-1}
 +\frac{1+c-ac-bc}{1-q}u\right)\right]
 \nonumber\\
 &=\frac{(abq^{d/2}/u;q)_\infty(q^{1+d/2}/u;q)_\infty
 (acq^{d/2}u;q)_\infty(bcq^{d/2}u;q)_\infty}
 {(aq^{d/2}/u;q)_\infty(bq^{d/2}/u;q)_\infty
 (q^{d/2}u;q)_\infty(cq^{d/2}u;q)_\infty}.
 \label{eq:S-Phi}
\end{align}
The fixed-sector index is then \cite{Kim:2009wb,Chang:2024lkw}
\begin{align}
 \I_{\ABJM,k}^{\{n_\alpha,N_\alpha\}}
 ={}&\frac{1}{\prod_\alpha(n_\alpha!)^2}
 \oint\prod_{\alpha,p}
 \frac{\dd z_{\alpha,p}}{2\pi iz_{\alpha,p}}
 \frac{\dd w_{\alpha,p}}{2\pi iw_{\alpha,p}}
 \nonumber\\[-2pt]
 &\times\prod_{(\alpha,p)\neq(\beta,p')}
 \left(1-q^{d_{\alpha\beta}/2}
 \frac{z_{\beta,p'}}{z_{\alpha,p}}\right)
 \left(1-q^{d_{\alpha\beta}/2}
 \frac{w_{\beta,p'}}{w_{\alpha,p}}\right)
 \nonumber\\[-2pt]
 &\times\prod_{\alpha,\beta}
 \prod_{p=1}^{n_\alpha}\prod_{p'=1}^{n_\beta}
 \Phi_{d_{\alpha\beta}}\!\left(
 \frac{z_{\alpha,p}}{w_{\beta,p'}}\right)
 \prod_{\alpha,p}z_{\alpha,p}^{-kN_\alpha}
 w_{\alpha,p}^{kN_\alpha}.
 \label{eq:S-abjm-fixed-index}
\end{align}
The last product in Eq.~\eqref{eq:S-abjm-fixed-index} is the right-hand side
of Eq.~\eqref{eq:S-CS-phase}; every other factor in
Eq.~\eqref{eq:S-abjm-fixed-index} is independent of $k$.  All holonomy
contours are oriented counterclockwise; the ABJM contours are nested
according to the prescription below.

\section{Contour prescription from the original holonomies}

The variables $Z_I$ and $W_I$ in Eq.~\eqref{eq:S-full-abjm-index} are
integrated on unit circles.  Equation~\eqref{eq:S-holonomy-rescale}
therefore sends that cycle to
\begin{equation}
 |z_{\alpha,p}|=r_z=|y_3|^{-1/2},
 \qquad
 |w_{\alpha,p}|=r_w=1.
 \label{eq:S-inherited-contours}
\end{equation}
We first define the integral in the real reference chamber
\begin{equation}
 a=b=c=t^2,
 \qquad q=t^6,
 \qquad y_3=t^{-1},
 \qquad 0<t<1,
 \label{eq:S-reference-chamber}
\end{equation}
for which
\begin{equation}
 r_z=t^{1/2}<r_w=1.
 \label{eq:S-reference-radii}
\end{equation}
For fixed $z$ and generic fugacities, the denominator of
$\Phi_d(z/w)$ in Eq.~\eqref{eq:S-Phi} gives the four pole families
\begin{align}
 w&=q^{d/2+\ell}z,
 &w&=cq^{d/2+\ell}z,
 &&\ell=0,1,2,\ldots,
 \label{eq:S-inner-pole-families}\\
 w&=\frac{z}{a q^{d/2+\ell}},
 &w&=\frac{z}{b q^{d/2+\ell}},
 &&\ell=0,1,2,\ldots.
 \label{eq:S-outer-pole-families}
\end{align}
The first two families lie inside the $w$ contour and the last two lie
outside whenever $|q|<1$ and
\begin{equation}
 \max(1,|c|)r_z<r_w<
 \min(|a|^{-1},|b|^{-1})r_z.
 \label{eq:S-contour-separation}
\end{equation}
At Eq.~\eqref{eq:S-reference-chamber}, the largest radius in
Eq.~\eqref{eq:S-inner-pole-families} is $t^{1/2}$ and the smallest
radius in Eq.~\eqref{eq:S-outer-pole-families} is $t^{-3/2}$, so the
unit $w$ circle separates the two sets strictly.

Equation~\eqref{eq:S-abjm-fixed-index} is defined first in the connected
reference chamber specified by Eqs.~\eqref{eq:S-inherited-contours} and
\eqref{eq:S-contour-separation}.  For other fugacity values, it is the
analytic continuation from this chamber with the cycles deformed
continuously so that the pole families in
Eq.~\eqref{eq:S-inner-pole-families} remain enclosed and those in
Eq.~\eqref{eq:S-outer-pole-families} remain excluded.  If a pole crosses
a fixed geometric circle, the circle is deformed with it; the continuation
does not change which of the two sets is enclosed.  The path is restricted
to avoid pinch loci at which an enclosed and an excluded pole collide.
This prescription preserves the homology class of the original
unit-circle cycle and also defines the index coefficientwise by expansion
about Eq.~\eqref{eq:S-reference-chamber}.

\section{Plane-wave pole selection}
\label{sec:S-plane-wave-poles}

We use the finite-family abbreviations of the Letter throughout.  With the
contour prescription above, the plane-wave contribution retains the
$d=0$, $\ell=0$ poles of the first family in
Eq.~\eqref{eq:S-inner-pole-families},
\begin{equation}
 w_{\alpha,p}=z_{\alpha,\sigma_\alpha(p)},
 \qquad \sigma_\alpha\in S_{n_\alpha}.
 \label{eq:S-permutation-poles}
\end{equation}
Since $d_{\alpha\alpha}=0$, the right vector determinant in
Eq.~\eqref{eq:S-abjm-fixed-index} contains the within-block factor
\begin{equation}
 \prod_\alpha\prod_{\substack{p,p'=1\\p\neq p'}}^{n_\alpha}
 \left(1-\frac{w_{\alpha,p'}}{w_{\alpha,p}}\right),
\end{equation}
which vanishes if two $w$ variables in the same block select the same
$z$ variable.  The contributing assignments are therefore one-to-one
within each block.

All permutations give the same contribution.  We fix the identity assignment $w_{\alpha,p}=z_{\alpha,p}$ and cancel one of the two Weyl factors $\prod_\alpha n_\alpha!$. The pole at $w=z$ equates the left and right holonomies, as expected from the breaking to the diagonal gauge group in Eq.~\eqref{eq:S-broken-gauge-group}.  At this assignment the last product in Eq.~\eqref{eq:S-abjm-fixed-index} is
\begin{equation}
 \left.\prod_{\alpha,p}z_{\alpha,p}^{-kN_\alpha}
 w_{\alpha,p}^{kN_\alpha}\right|_{w_{\alpha,p}=z_{\alpha,p}}
 =\prod_{\alpha,p}z_{\alpha,p}^{-kN_\alpha}
 z_{\alpha,p}^{kN_\alpha}=1.
 \label{eq:S-CS-factor-at-selected-pole}
\end{equation}
\subsection{Coefficientwise selection at finite momentum}
\label{sec:S-fixed-grade-selection}

The full index agrees with the selected residues below an explicit grade:
for $r_i\in\frac12\mathbb Z_{\geq0}$,
\begin{equation}
 [a^{r_1}b^{r_2}c^{r_3}]
 \left(\I_{\ABJM,k;\Lambda}
       -\I_{\ABJM,k;\Lambda}^{\rm pw}\right)=0,
 \qquad
 r_1+r_2+r_3<1+k\min_\alpha N_\alpha.
 \label{eq:S-pole-error}
\end{equation}
Here brackets extract a monomial coefficient in the formal expansion
specified by the contour prescription.  This statement holds for every
finite set of distinct positive $N_\alpha$ and positive multiplicities
$n_\alpha$, at each positive integer level $k$.

To prove it, isolate the zero-grade matter factor
\begin{equation}
 \mathcal C(z,w)=
 \prod_\alpha\prod_{p,p'=1}^{n_\alpha}
 \left(1-\frac{z_{\alpha,p}}{w_{\alpha,p'}}\right)^{-1}.
 \label{eq:S-zero-grade-Cauchy}
\end{equation}
Let $\mathcal F(z,w;a,b,c)$ be the remaining one-loop factor after
removing $\mathcal C(z,w)$ and both within-block vector determinants
$\prod_\alpha\Delta(z_\alpha)\Delta(w_\alpha)$ from
Eq.~\eqref{eq:S-abjm-fixed-index}.  Equivalently, replace each within-block
$\Phi_0(u)$ by $(1-u)\Phi_0(u)$ and retain all cross-block vector factors.
The factor $\mathcal C$ is expanded with $|z_{\alpha,p}|<|w_{\alpha,p'}|$.
The other factors are expanded in nonnegative half-powers of $a,b,c$.
For sufficiently small fugacities these expansions converge on fixed
nested circles satisfying Eq.~\eqref{eq:S-contour-separation}, so
coefficient extraction may be performed before contour integration.
Every remaining elementary factor in Eq.~\eqref{eq:S-Phi} and the vector
determinants has the form $(1-Mv)^{\pm1}$, where $M$ is a fugacity monomial
of total grade at least one and $v$ is a ratio of two holonomies.
In particular, cross-block vector factors have grade
$3d_{\alpha\beta}/2\geq3/2$, since the block sizes are distinct.

Fix $\mathbf r=(r_1,r_2,r_3)$ and write
$|\mathbf r|=r_1+r_2+r_3$ and
$\mathcal F_{\mathbf r}=[a^{r_1}b^{r_2}c^{r_3}]\mathcal F$.
Only finitely many factors of the Pochhammer products and finitely many
terms of their geometric expansions can contribute to this coefficient.
Thus $\mathcal F_{\mathbf r}$ is a finite Laurent polynomial, symmetric
separately in each set of block holonomies.  A monomial contributing to
it contains at most $\lfloor|\mathbf r|\rfloor$ positive-grade factors,
counted with multiplicity.  Its exponent in any individual $z$ or $w$
therefore lies between $-\lfloor|\mathbf r|\rfloor$ and
$\lfloor|\mathbf r|\rfloor$.  This bound controls the complete one-loop
coefficient before any pole is substituted.

Consider one $w_{\alpha,p}$ contour at this fixed grade, keeping the other
holonomies nonzero and generic.  The factor $\mathcal C$ has a zero of
order $n_\alpha$ at $w_{\alpha,p}=0$, whereas
$\Delta(w_\alpha)$ has a pole of order at most $n_\alpha-1$.
Including the Chern--Simons power and the measure, the part of the
integrand that can be singular at the origin obeys
\begin{align}
 &w_{\alpha,p}^{kN_\alpha-1}\Delta(w_\alpha)
 \prod_{p'=1}^{n_\alpha}
 \left(1-\frac{z_{\alpha,p'}}{w_{\alpha,p}}\right)^{-1}
 \mathcal F_{\mathbf r}(z,w)
 \nonumber\\[-2pt]
 &\hspace{35mm}
 =O\!\left(w_{\alpha,p}^{kN_\alpha-\lfloor|\mathbf r|\rfloor}\right).
 \label{eq:S-fixed-grade-origin}
\end{align}
Consequently the origin is regular whenever
$kN_\alpha\geq\lfloor|\mathbf r|\rfloor$.
At fixed grade the integrand is rational in the $w$ variables, and its
only other poles are $w_{\alpha,p}=z_{\alpha,p'}$ within the same block.
The same bound applies successively to every $w$ integral.

Only permutation assignments survive the within-block vector zeros.
For any such assignment, the non-pole factors of $\mathcal C$ cancel
$\prod_\alpha\Delta(w_\alpha)$ at the residue, while the product of
Chern--Simons monomials is one.  Summing the permutations cancels one
Weyl factor.  Hence, if
$k\min_\alpha N_\alpha\geq\lfloor|\mathbf r|\rfloor$,
\begin{align}
 [a^{r_1}b^{r_2}c^{r_3}]\I_{\ABJM,k;\Lambda}
 &=\frac{1}{\prod_\alpha n_\alpha!}
 \oint\prod_{\alpha,p}\frac{\dd z_{\alpha,p}}{2\pi i z_{\alpha,p}}
 \prod_\alpha\Delta(z_\alpha)\,
 \mathcal F_{\mathbf r}(z,z)
 \nonumber\\[-2pt]
 &=[a^{r_1}b^{r_2}c^{r_3}]
 \I_{\ABJM,k;\Lambda}^{\rm pw}.
 \label{eq:S-fixed-grade-reduction}
\end{align}
This proves Eq.~\eqref{eq:S-pole-error}, including the remaining
$z$-holonomy projection.  For any fixed total-grade cutoff $D$, all
coefficients through that grade agree once
$k(\Lambda+\min_\alpha\nu_\alpha)\geq\lfloor D\rfloor$.
The expansion at fixed grade reduces the contour calculation to finitely
many poles, so the proof requires no exchange of the large-$\Lambda$
limit with the infinite pole families accumulating at $w=0$.
Equation~\eqref{eq:S-pole-error} is a formal coefficientwise statement;
it does not assert uniform analytic convergence in the fugacities.

\subsection{Evaluation of the selected Higgs residues}

We now reduce the selected residues to the one-holonomy integral in
Eq.~\eqref{eq:S-common-integral}.

With $N_\alpha=\Lambda+\nu_\alpha$ and fixed $\nu_\alpha$, Eq.~\eqref{eq:S-matrix-kernel} gives
\begin{align}
 H^{\rm mono}_{\alpha\beta}(x)
 &\equiv
 \lim_{\Lambda\to\infty}H^{\rm mat}_{\alpha\beta}(x)
 \nonumber\\[-2pt]
 &=(1-x)^{-\delta_{\alpha\beta}}
 \prod_{\ell=0}^{\infty}
 \mathcal K\!\left((-1)^{d_{\alpha\beta}}
 q^{d_{\alpha\beta}/2+\ell}x\right).
 \label{eq:S-monopole-kernel}
\end{align}
The following calculation identifies their remaining integrand with
$\I[H^{\rm mono}]$.

We begin with the simple pole responsible for each individual $w$ residue.  Setting $d=0$ in Eq.~\eqref{eq:S-Phi} gives
\begin{equation}
 \Phi_0(x)=
 \frac{(ab/x;q)_\infty(q/x;q)_\infty
 (acx;q)_\infty(bcx;q)_\infty}
 {(a/x;q)_\infty(b/x;q)_\infty
 (x;q)_\infty(cx;q)_\infty}.
 \label{eq:S-Phi-zero}
\end{equation}
The only factor singular at $x=1$ is $(x;q)_\infty^{-1}$.  Using
\begin{equation}
 (x;q)_\infty=(1-x)(qx;q)_\infty,
 \label{eq:S-Pochhammer-shift-repeated}
\end{equation}
we find
\begin{align}
 (1-x)\Phi_0(x)
 ={}&
 \frac{(ab/x;q)_\infty(q/x;q)_\infty
 (acx;q)_\infty(bcx;q)_\infty}
 {(a/x;q)_\infty(b/x;q)_\infty
 (qx;q)_\infty(cx;q)_\infty}.
 \label{eq:S-diagonal-pole-step}
\end{align}
Taking $x\to1$, the two factors $(q;q)_\infty$ cancel, so
\begin{equation}
 \lim_{x\to1}(1-x)\Phi_0(x)
 =
 \frac{(ab;q)_\infty(ac;q)_\infty(bc;q)_\infty}
 {(a;q)_\infty(b;q)_\infty(c;q)_\infty}.
 \label{eq:S-diagonal-pole-value}
\end{equation}
For $\alpha=\beta$, Eq.~\eqref{eq:S-monopole-kernel} gives
\begin{align}
 H^{\rm mono}_{\alpha\alpha}(x)
 &=\frac{1}{1-x}
 \prod_{\ell=0}^{\infty}\mathcal K(q^\ell x)
 \nonumber\\
 &=
 \frac{(abx;q)_\infty(acx;q)_\infty(bcx;q)_\infty}
 {(ax;q)_\infty(bx;q)_\infty(cx;q)_\infty}.
 \label{eq:S-hmat-diagonal}
\end{align}
Here the factor $1-x$ in $\mathcal K(x)$ is canceled against the displayed
$1-x$ in the denominator before setting $x=1$.  Hence
\begin{equation}
 H^{\rm mono}_{\alpha\alpha}(1)
 =\frac{(ab;q)_\infty(ac;q)_\infty(bc;q)_\infty}
 {(a;q)_\infty(b;q)_\infty(c;q)_\infty}.
 \label{eq:S-monopole-kernel-diagonal}
\end{equation}
Consequently,
\begin{equation}
 {
 \lim_{x\to1}(1-x)\Phi_0(x)
 =H^{\rm mono}_{\alpha\alpha}(1).}
 \label{eq:S-Phi-h-diagonal}
\end{equation}
This is the residue needed for one diagonal pair $(z_{\alpha,p},w_{\alpha,p})$.  Set $x=z_{\alpha,p}/w_{\alpha,p}$.  Near $w_{\alpha,p}=z_{\alpha,p}$,
\begin{equation}
 1-x=1-\frac{z_{\alpha,p}}{w_{\alpha,p}}
 =\frac{w_{\alpha,p}-z_{\alpha,p}}{w_{\alpha,p}}.
 \label{eq:S-local-residue-coordinate}
\end{equation}
Therefore
\begin{align}
 \Res_{w_{\alpha,p}=z_{\alpha,p}}
 \left[
 \frac{\dd w_{\alpha,p}}{w_{\alpha,p}}
 \Phi_0\!\left(\frac{z_{\alpha,p}}{w_{\alpha,p}}\right)
 \right]
 =H^{\rm mono}_{\alpha\alpha}(1).
 \label{eq:S-one-w-residue}
\end{align}

We next prove the identity that combines the remaining matter factors with the two vector determinants.  For any pair of blocks $\alpha,\beta$, direct multiplication of Eq.~\eqref{eq:S-Phi} and its $x\mapsto x^{-1}$ image gives
\begin{align}
 &\left[
 \left(1-q^{d_{\alpha\beta}/2}x\right)
 \left(1-q^{d_{\alpha\beta}/2}x^{-1}\right)
 \right]^2
 \Phi_{d_{\alpha\beta}}(x)
 \Phi_{d_{\alpha\beta}}(x^{-1})
 \nonumber\\
 &\quad=
 \left(1-q^{d_{\alpha\beta}/2}x\right)
 \left(1-q^{d_{\alpha\beta}/2}x^{-1}\right)
 \nonumber\\[-2pt]
 &\qquad\times
 \frac{(abq^{d_{\alpha\beta}/2}x;q)_\infty
 (acq^{d_{\alpha\beta}/2}x;q)_\infty
 (bcq^{d_{\alpha\beta}/2}x;q)_\infty}
 {(aq^{d_{\alpha\beta}/2}x;q)_\infty
 (bq^{d_{\alpha\beta}/2}x;q)_\infty
 (cq^{d_{\alpha\beta}/2}x;q)_\infty}
 \nonumber\\[-2pt]
 &\qquad\times
 \frac{(abq^{d_{\alpha\beta}/2}/x;q)_\infty
 (acq^{d_{\alpha\beta}/2}/x;q)_\infty
 (bcq^{d_{\alpha\beta}/2}/x;q)_\infty}
 {(aq^{d_{\alpha\beta}/2}/x;q)_\infty
 (bq^{d_{\alpha\beta}/2}/x;q)_\infty
 (cq^{d_{\alpha\beta}/2}/x;q)_\infty}.
 \label{eq:S-Phi-pair-expanded}
\end{align}
The reduction from the first line to the remaining lines uses Eq.~\eqref{eq:S-Pochhammer-shift-repeated} twice.  For example,
\begin{equation}
 \frac{(q^{1+d/2}x;q)_\infty}{(q^{d/2}x;q)_\infty}
 =\frac{1}{1-q^{d/2}x},
 \label{eq:S-shift-example}
\end{equation}
with $d=d_{\alpha\beta}$, and similarly for $x^{-1}$.  The right-hand side can now be recognized from the limiting BMN kernel.  Multiplying the infinite product in Eq.~\eqref{eq:S-monopole-kernel} gives
\begin{align}
 &H^{\rm mono}_{\alpha\beta}
 \!\left((-1)^{d_{\alpha\beta}}x\right)
 \nonumber\\
 &\quad=
 \frac{1-q^{d_{\alpha\beta}/2}x}
 {\left(1-(-1)^{d_{\alpha\beta}}x\right)^{\delta_{\alpha\beta}}}
 \frac{(abq^{d_{\alpha\beta}/2}x;q)_\infty
 (acq^{d_{\alpha\beta}/2}x;q)_\infty
 (bcq^{d_{\alpha\beta}/2}x;q)_\infty}
 {(aq^{d_{\alpha\beta}/2}x;q)_\infty
 (bq^{d_{\alpha\beta}/2}x;q)_\infty
 (cq^{d_{\alpha\beta}/2}x;q)_\infty}.
 \label{eq:S-hmat-Pochhammer}
\end{align}
Here the factor $1-q^{d_{\alpha\beta}/2}x$ comes from the ratio
$(q^{d_{\alpha\beta}/2}x;q)_\infty/
(q^{1+d_{\alpha\beta}/2}x;q)_\infty$.
Multiplying Eq.~\eqref{eq:S-hmat-Pochhammer} by its $x\mapsto x^{-1}$ counterpart and comparing with Eq.~\eqref{eq:S-Phi-pair-expanded} proves
\begin{align}
 &{
 \left[
 \left(1-q^{d_{\alpha\beta}/2}x\right)
 \left(1-q^{d_{\alpha\beta}/2}x^{-1}\right)
 \right]^2
 \Phi_{d_{\alpha\beta}}(x)
 \Phi_{d_{\alpha\beta}}(x^{-1})}
 \nonumber\\[-2pt]
 &{
 \qquad=
 \left[(1-x)(1-x^{-1})\right]^{\delta_{\alpha\beta}}
 H^{\rm mono}_{\alpha\beta}
 \!\left((-1)^{d_{\alpha\beta}}x\right)
 H^{\rm mono}_{\beta\alpha}
 \!\left((-1)^{d_{\alpha\beta}}x^{-1}\right).}
 \label{eq:S-Phi-h-pair}
\end{align}
For $\alpha=\beta$, the factor $(1-x)(1-x^{-1})$ on the second line is one of the factors in $\Delta(u_\alpha)$ in Eq.~\eqref{eq:S-haar}.  For $\alpha\neq\beta$, $\delta_{\alpha\beta}=0$, so no such factor remains between different blocks.

We can now carry out the $w$ integrations. Consider first the $n_\alpha$ matter factors whose poles are taken,
\begin{equation}
 \prod_{p=1}^{n_\alpha}
 \Phi_0\!\left(\frac{z_{\alpha,p}}{w_{\alpha,p}}\right).
 \label{eq:S-selected-diagonal-factors}
\end{equation}
Equation~\eqref{eq:S-one-w-residue} shows that their residues give
\begin{equation}
 \prod_{p=1}^{n_\alpha}H^{\rm mono}_{\alpha\alpha}(1).
 \label{eq:S-diagonal-residue-product}
\end{equation}
Next choose two distinct eigenvalues $z_{\alpha,p'},z_{\alpha,p}$ with $p\neq p'$ in the same block $\alpha$.  Before taking the residue, this unordered pair appears once in each orientation. After setting $w=z$, its complete contribution is
\begin{align}
 &\left[\left(1-\frac{z_{\alpha,p'}}{z_{\alpha,p}}\right)\left(1-\frac{z_{\alpha,p}}{z_{\alpha,p'}}\right)\right]^2
 \Phi_0\left(\frac{z_{\alpha,p'}}{z_{\alpha,p}}\right)\Phi_0\left(\frac{z_{\alpha,p}}{z_{\alpha,p'}}\right)
 \nonumber\\
 &\qquad=\left(1-\frac{z_{\alpha,p'}}{z_{\alpha,p}}\right)\left(1-\frac{z_{\alpha,p}}{z_{\alpha,p'}}\right)
 H^{\rm mono}_{\alpha\alpha}\left(\frac{z_{\alpha,p'}}{z_{\alpha,p}}\right)
 H^{\rm mono}_{\alpha\alpha}\left(\frac{z_{\alpha,p}}{z_{\alpha,p'}}\right),
 \label{eq:S-same-block-pair}
\end{align}
where Eq.~\eqref{eq:S-Phi-h-pair} was used with $d_{\alpha\alpha}=0$.  Taking the product over all unordered pairs $p,p'$ gives
\begin{equation}
 \Delta(z_\alpha)
 \prod_{p\neq p'}
 H^{\rm mono}_{\alpha\alpha}
 \!\left(\frac{z_{\alpha,p'}}{z_{\alpha,p}}\right).
 \label{eq:S-same-block-all-pairs}
\end{equation}
Combining this with Eq.~\eqref{eq:S-diagonal-residue-product} fills in the missing $p=p'$ terms and yields
\begin{equation}
 \Delta(z_\alpha)
 \prod_{p,p'=1}^{n_\alpha}
 H^{\rm mono}_{\alpha\alpha}
 \!\left(\frac{z_{\alpha,p'}}{z_{\alpha,p}}\right).
 \label{eq:S-complete-diagonal-block}
\end{equation}
For two different blocks $\alpha\neq\beta$, there is again one factor in each orientation.  Since now $\delta_{\alpha\beta}=0$, Eq.~\eqref{eq:S-Phi-h-pair} gives
\begin{align}
 &H^{\rm mono}_{\alpha\beta}
 \!\left((-1)^{d_{\alpha\beta}}
 \frac{z_{\beta,p'}}{z_{\alpha,p}}\right)
 H^{\rm mono}_{\beta\alpha}
 \!\left((-1)^{d_{\alpha\beta}}
 \frac{z_{\alpha,p}}{z_{\beta,p'}}\right).
 \label{eq:S-cross-block-pair}
\end{align}
This accounts for the complete unordered pair of cross-block eigenvalues.

Finally set
\begin{equation}
 u_{\alpha,p}=(-1)^{N_\alpha}z_{\alpha,p}.
 \label{eq:S-holonomy-redefinition}
\end{equation}
The measure and the within-block Vandermonde are unchanged because all variables in a fixed block are multiplied by the same sign.  Moreover,
\begin{equation}
 \frac{u_{\beta,p'}}{u_{\alpha,p}}
 =(-1)^{N_\beta-N_\alpha}
 \frac{z_{\beta,p'}}{z_{\alpha,p}}
 =(-1)^{d_{\alpha\beta}}
 \frac{z_{\beta,p'}}{z_{\alpha,p}},
 \label{eq:S-sign-ratio}
\end{equation}
where $N_\beta-N_\alpha$ and $|N_\beta-N_\alpha|$ have the same parity.
The resulting integrand depends only on holonomy ratios.  Its common
radius can therefore be changed from $r_z$ to one without changing any
ratio or crossing a pole.  Therefore Eqs.~\eqref{eq:S-complete-diagonal-block}
and \eqref{eq:S-cross-block-pair}, together with the permutation sum, give
\begin{align}
 \I_{\ABJM,k;\Lambda}^{\rm pw}
 ={}&\frac{1}{\prod_\alpha n_\alpha!}
 \oint\prod_{\alpha,p}\frac{\dd u_{\alpha,p}}{2\pi i u_{\alpha,p}}
 \prod_\alpha\Delta(u_\alpha)
 \nonumber\\[-2pt]
 &\times\prod_{\alpha,\beta}
 \prod_{p=1}^{n_\alpha}\prod_{p'=1}^{n_\beta}
 H^{\rm mono}_{\alpha\beta}
 \!\left(\frac{u_{\beta,p'}}{u_{\alpha,p}}\right)
 \nonumber\\
 ={}&\I[H^{\rm mono}].
 \label{eq:S-abjm-plane-wave-common}
\end{align}
This is exactly the form of Eq.~\eqref{eq:S-common-integral}: on the selected residues, the $w$ variables have been eliminated, one ABJM Weyl factor has been canceled by the permutation sum, and the two original gauge holonomies have become the single diagonal holonomy $u$.

This proves the level-independent common-integral identification for the
plane-wave contribution.  Equation~\eqref{eq:S-pole-error} controls its
difference from the full two-holonomy index at finite momentum.

\section{Fusion coefficients}
\label{sec:S-fusion}

The Letter compares the continuum monopole-harmonic and finite matrix-harmonic ranges.  Here we supply the additional comparison of their product algebras.  Let
$\widetilde Y^{(N_\alpha-N_\beta)/2}_{jm}$ be normalized scalar monopole harmonics and let
$\widehat Y^{(\alpha\beta)}_{jm}\in
\Hom(\mathbb C^{N_\beta},\mathbb C^{N_\alpha})$ be normalized rectangular matrix harmonics.  In a standard phase convention, the continuum fusion coefficient is
\begin{align}
 &\int_{S^2}\frac{\dd\Omega}{4\pi}
 \left(\widetilde Y^{(N_\alpha-N_\gamma)/2}_{j_1m_1}\right)^*
 \widetilde Y^{(N_\alpha-N_\beta)/2}_{j_2m_2}
 \widetilde Y^{(N_\beta-N_\gamma)/2}_{j_3m_3}
 \nonumber\\[-2pt]
 &\quad=
 \sqrt{\frac{(2j_2+1)(2j_3+1)}{2j_1+1}}
 C^{j_1m_1}_{j_2m_2\,j_3m_3}
 C^{j_1,(N_\alpha-N_\gamma)/2}_{
 j_2,(N_\alpha-N_\beta)/2\,
 j_3,(N_\beta-N_\gamma)/2}.
 \label{eq:S-continuum-fusion}
\end{align}
Each Clebsch--Gordan coefficient can equivalently be written as a Wigner $3j$ symbol,
\begin{equation}
 C^{j_1m_1}_{j_2m_2\,j_3m_3}
 =(-1)^{j_2-j_3+m_1}\sqrt{2j_1+1}
 \begin{pmatrix}
 j_2&j_3&j_1\\
 m_2&m_3&-m_1
 \end{pmatrix}.
 \label{eq:S-CG-threej}
\end{equation}
The finite matrix coefficient is
\begin{align}
 &\frac{1}{\Lambda}\Tr\left[
 \left(\widehat Y^{(\alpha\gamma)}_{j_1m_1}\right)^\dagger
 \widehat Y^{(\alpha\beta)}_{j_2m_2}
 \widehat Y^{(\beta\gamma)}_{j_3m_3}\right]
 \nonumber\\[-2pt]
 &\quad=(-1)^{j_1+(N_\gamma-1)/2+(N_\alpha-1)/2}
 C^{j_1m_1}_{j_2m_2\,j_3m_3}
 \nonumber\\[-2pt]
 &\qquad\times\sqrt{\Lambda(2j_2+1)(2j_3+1)}
 \begin{Bmatrix}
 j_1&j_2&j_3\\
 (N_\beta-1)/2&(N_\gamma-1)/2&(N_\alpha-1)/2
 \end{Bmatrix},
 \label{eq:S-matrix-fusion}
\end{align}
where $N_\alpha=\Lambda+\nu_\alpha$.  At fixed $j_i$ and fixed differences $N_\alpha-N_\beta$, the Wigner $6j$ symbol obeys \cite{flude1997edmonds}
\begin{align}
 &(-1)^{j_1+(N_\gamma-1)/2+(N_\alpha-1)/2}
 \sqrt{\Lambda(2j_1+1)}
 \begin{Bmatrix}
 j_1&j_2&j_3\\
 (N_\beta-1)/2&(N_\gamma-1)/2&(N_\alpha-1)/2
 \end{Bmatrix}
 \nonumber\\[-2pt]
 &\hspace{25mm}\longrightarrow
 C^{j_1,(N_\alpha-N_\gamma)/2}_{
 j_2,(N_\alpha-N_\beta)/2\,
 j_3,(N_\beta-N_\gamma)/2}.
 \label{eq:S-sixj-limit}
\end{align}
Substitution into Eq.~\eqref{eq:S-matrix-fusion} reproduces Eq.~\eqref{eq:S-continuum-fusion}.  Thus the large common momentum removes both the upper harmonic cutoff and the finite-matrix deformation of the fusion coefficients.

\section{Low-rank checks by constant-term extraction}
\label{sec:S-low-rank-check}

Consider two sectors with $n_1=n_2=1$ and
$(N_1,N_2)=(3,4)$ or $(4,5)$.
Their BMN ranks are $\mathcal N=7$ and $9$, respectively; both have
ABJM gauge rank $r=2$ and
$\vec m=\widetilde{\vec m}=(-N_1,-N_2)$.
Set $a=b=c=t^2$ and $q=t^6$, with $0<t<1$.

The common shift $(3,4)\to(4,5)$ keeps $d_{12}=1$ and therefore leaves
the plane-wave index unchanged.  The theoretical onsets of the two
corrections in Eq.~(4) of the Letter follow from the cutoff and pole
factors.  Comparing Eqs.~\eqref{eq:S-matrix-kernel} and
\eqref{eq:S-monopole-kernel}, the earliest omitted factor comes from
the smaller diagonal block:
$\mathcal K(q^{N_1})=1-t^{6N_1}+O(t^{6N_1+2})$.
All other omitted factors start at higher degree, so the harmonic tail
is predicted to begin with $-t^{6N_1}$.

For the non-plane-wave correction, the lowest competing diagonal pole
in Eq.~\eqref{eq:S-inner-pole-families} is
$w_1=cz_1$, with $w_2=z_2$.
Its Chern--Simons weight is
\begin{equation}
\prod_{\alpha,p}z_{\alpha,p}^{-kN_\alpha}
 w_{\alpha,p}^{kN_\alpha} \bigg|_{w_1=cz_1,~w_2=z_2} = c^{k N_1}.
 \label{97eq}
\end{equation}
After taking the pole from $(1-cz_1/w_1)^{-1}$ in
Eq.~\eqref{eq:S-Phi-zero}, the remaining factor gives
\begin{equation*}
 \left.(1-z_1/w_1)^{-1}\right|_{w_1=cz_1}
 =-\frac{c}{1-c}=-t^2+O(t^4).
\end{equation*}
All other factors and the remaining holonomy integral have leading
term one.  This residue therefore predicts the leading correction
$-t^{2kN_1+2}$.  The analogous pole in the larger block starts at
$t^{2kN_2+2}$.
Every other nontrivial pole ratio has $t$-degree at least three;
power counting the complete one-loop prefactors in
Eq.~\eqref{eq:S-abjm-fixed-index} gives nonnegative degree.
For the sectors and levels considered here,
$3kN_1>2kN_1+2$, so these residues cannot contribute earlier.

Thus the predicted harmonic-tail onsets are $t^{18}$ and $t^{24}$
for $(3,4)$ and $(4,5)$, respectively.
The predicted non-plane-wave corrections start at $t^8$ and $t^{10}$ for $k=1$,
and $t^{14}$ and $t^{18}$ for $k=2$; all predicted leading coefficients
are $-1$.
We expand the two sectors through $t^{18}$ and $t^{24}$ to test both
predicted corrections.

We use the constant-term extraction method described in Sec.~4.4 of
Ref.~\cite{Chang:2026mui}.  We first expand the integrand with respect to $t$ to a finite order before integration,
keeping the holonomies as Laurent monomials. For the given finite order of $t$, only finitely many factors in the integrand contribute, as shown later. The contour integrals then
select the terms with zero power of every holonomy, since
\begin{equation}
 \oint\frac{\dd x}{2\pi i x}\,x^\ell=\delta_{\ell,0}.
 \label{eq:S-low-rank-CT}
\end{equation}
We denote this projection by $[x^0]$.  

On the BMN side, write $u_\alpha=u_{\alpha,1}$ and
$x=u_2/u_1$.  The overall holonomy decouples, and
Eq.~\eqref{eq:S-common-integral} reduces to
\begin{equation}
 \I[H]=[x^0]\left[
 H_{11}(1)H_{22}(1)H_{12}(x)H_{21}(x^{-1})\right].
 \label{eq:S-two-block-kernel-CT}
\end{equation}
Inserting $H^{\rm mat}$ from Eq.~\eqref{eq:S-matrix-kernel} gives the
finite BMN index; inserting $H^{\rm mono}$ from
Eq.~\eqref{eq:S-monopole-kernel} gives the ABJM plane-wave contribution.
The removable diagonal factors are canceled before expansion, as in
Eq.~\eqref{eq:S-matrix-kernel-diagonal}.  The overall $U(1)$ multiplet
contribution is retained.  All Weyl factors are one for $n_1=n_2=1$.

For the full ABJM sector, abbreviate $z_\alpha=z_{\alpha,1}$ and
$w_\alpha=w_{\alpha,1}$.  Equation~\eqref{eq:S-abjm-fixed-index} becomes
\begin{align}
 \I_{\ABJM,k}^{\{(1,N_1),(1,N_2)\}}
 ={}&[z_1^0z_2^0w_1^0w_2^0]\Bigg[
 \left(\frac{w_1}{z_1}\right)^{kN_1}
 \left(\frac{w_2}{z_2}\right)^{kN_2}
 \nonumber\\
 &\times\prod_{\alpha\ne\beta}
 \left(1-t^3\frac{z_\beta}{z_\alpha}\right)
 \left(1-t^3\frac{w_\beta}{w_\alpha}\right)
 \nonumber\\
 &\times\Phi_0\!\left(\frac{z_1}{w_1}\right)
 \Phi_0\!\left(\frac{z_2}{w_2}\right)
 \Phi_1\!\left(\frac{z_1}{w_2}\right)
 \Phi_1\!\left(\frac{z_2}{w_1}\right)\Bigg],
 \label{eq:S-two-block-abjm-CT}
\end{align}
where $\Phi_d$ is evaluated at $a=b=c=t^2$ in Eq.~\eqref{eq:S-Phi}.
The Laurent expansion is fixed by the contour chamber
$|z_1|=|z_2|=t^{1/2}$ and $|w_1|=|w_2|=1$.
Each denominator factor is expanded geometrically in this chamber;
in particular, the two factors of $t$-degree zero are expanded as
\begin{equation}
 \frac{1}{1-z_\alpha/w_\alpha}
 =\sum_{h_\alpha\geq0}
 \left(\frac{z_\alpha}{w_\alpha}\right)^{h_\alpha},
 \qquad \alpha=1,2.
 \label{eq:S-two-block-zero-mode}
\end{equation}
Their truncation is fixed by holonomy charge balance.  The classical
Chern--Simons monomial in the first line on the RHS of Eq.~\eqref{eq:S-two-block-abjm-CT} has total $z$-power $-k(N_1+N_2)$, and each inverse matter
ratio $w_\beta/z_\alpha$ in the last line costs at least $t^2$.  Consequently,
$h_1+h_2\leq7k+9$ suffices through $t^{18}$ for $(N_1,N_2)=(3,4)$,
and $h_1+h_2\leq9k+12$ suffices through $t^{24}$ for $(4,5)$.
All remaining factors have finite expansions at the chosen order.
After multiplying them, including the Chern--Simons monomial, we keep
only the coefficient of $z_1^0z_2^0w_1^0w_2^0$.

The constant-term results in Tables~\ref{tab:S-two-block-check}
and~\ref{tab:S-shifted-two-block-check} test these predicted leading corrections
and coefficients.

\begin{table}[ht]
 \caption{Coefficients of $t^j$ for
 $(n_1,N_1)=(1,3)$ and $(n_2,N_2)=(1,4)$.
 The common sector label is suppressed in the column headings.
 All odd-power coefficients through $t^{18}$ vanish.}
 \label{tab:S-two-block-check}
 \begin{ruledtabular}
 \begin{tabular}{rcccc}
 $j$ & $\I_{\BMN}$ & $\I_{\ABJM}^{\rm pw}$
 & $\I_{\ABJM,1}$ & $\I_{\ABJM,2}$\\
 \hline
 0  & 1  & 1  & 1  & 1\\
 2  & 6  & 6  & 6  & 6\\
 4  & 15 & 15 & 15 & 15\\
 6  & 21 & 21 & 21 & 21\\
 8  & 21 & 21 & 20 & 21\\
 10 & 30 & 30 & 27 & 30\\
 12 & 57 & 57 & 52 & 57\\
 14 & 57 & 57 & 50 & 56\\
 16 & 12 & 12 & 3  & 10\\
 18 & 38 & 37 & 27 & 33
 \end{tabular}
 \end{ruledtabular}
\end{table}

\clearpage
\begin{table}[ht]
 \caption{Coefficients of $t^j$ for
 $(n_1,N_1)=(1,4)$ and $(n_2,N_2)=(1,5)$,
 with the same column conventions as Table~\ref{tab:S-two-block-check}.
 All odd-power coefficients through $t^{24}$ vanish.}
 \label{tab:S-shifted-two-block-check}
 \begin{ruledtabular}
 \begin{tabular}{rcccc}
 $j$ & $\I_{\BMN}$ & $\I_{\ABJM}^{\rm pw}$
 & $\I_{\ABJM,1}$ & $\I_{\ABJM,2}$\\
 \hline
 0  & 1   & 1   & 1   & 1\\
 2  & 6   & 6   & 6   & 6\\
 4  & 15  & 15  & 15  & 15\\
 6  & 21  & 21  & 21  & 21\\
 8  & 21  & 21  & 21  & 21\\
 10 & 30  & 30  & 29  & 30\\
 12 & 57  & 57  & 54  & 57\\
 14 & 57  & 57  & 52  & 57\\
 16 & 12  & 12  & 5   & 12\\
 18 & 37  & 37  & 28  & 36\\
 20 & 150 & 150 & 139 & 148\\
 22 & 108 & 108 & 96  & 104\\
 24 & -78 & -79 & -95 & -85
 \end{tabular}
 \end{ruledtabular}
\end{table}

\bibliography{references}